\documentclass[preprint]{elsarticle}

\usepackage[utf8]{inputenc}
\usepackage[colorlinks,bookmarksopen,bookmarksnumbered,citecolor=black,urlcolor=black,linkcolor=black]{hyperref}
\usepackage{graphicx}
\usepackage{listings, lstautogobble}
\usepackage{xcolor}
\usepackage[T1]{fontenc}
\usepackage{amssymb}
\usepackage[ruled, linesnumbered]{algorithm2e}
\usepackage{caption}
\usepackage{balance}
\usepackage{enumitem}
\usepackage[normalem]{ulem}
\definecolor{codebackground}{rgb}{1,1,1}
\definecolor{black}{rgb}{0.13, 0.55, 0.13}
\usepackage{svg}
\usepackage{amsmath}
\usepackage{wrapfig}
\usepackage{lscape}
\usepackage{rotating}
\usepackage{epstopdf}
\usepackage{hologo} 
\usepackage{booktabs} 
\usepackage{threeparttable} 
\usepackage{nomencl}
\usepackage{shellesc}
\makenomenclature

\begin{document}
        \title{Towards an Extensible Model-Based Digital Twin Framework for Space Launch Vehicles}
        \author[lan,dut]{Ran Wei}
        \ead{r.wei5@lancaster.ac.uk;ranwei@dlut.edu.cn}
		
		\author[dut]{Ruizhe Yang\corref{mycorrespondingauthor}}
		\ead{ruizheyang@mail.dlut.edu.cn}

        \author[dut]{Shijun Liu\corref{mycorrespondingauthor}}
		\ead{liushijun@mail.dlut.edu.cn}

        \author[sae]{Chongsheng Fan}

        \author[sae]{Rong Zhou}
  
        \author[dut]{Zekun Wu}

        \author[dut]{Haochi Wang}

        \author[dut]{Yifan Cai}

        \author[seu]{Zhe Jiang\corref{mycorrespondingauthor}}
        \ead{101013615@seu.edu.cn}
		
        \address[lan]{Lancaster University, Lancaster, UK}
        \address[dut]{Dalian University of Technology, Dalian, China}
        \address[sae]{Shanghai Aerospace Equipment Manufacturing Co., Ltd, Shanghai, China}
        \address[seu]{Southeast University, Nanjing, China}
		
		\cortext[mycorrespondingauthor]{Corresponding authors}

\begin{abstract}
The concept of Digital Twin (DT) is increasingly applied to systems on different levels of abstraction across domains, to support monitoring, analysis, diagnosis, decision making and automated control. 
Whilst the interest in applying DT is growing, the definition of DT is unclear, neither is there a clear pathway to develop DT to fully realise its capacities.
In this paper, we revise the concept of DT and its categorisation.
We propose a DT maturity matrix, based on which we propose a model-based DT development methodology.
We also discuss how model-based tools can be used to support the methodology and present our own supporting tool.
We report our preliminary findings with a discussion on a case study, in which we use our proposed methodology and our supporting tool to develop an extensible DT platform for the assurance of Electrical and Electronics systems of space launch vehicles.
\end{abstract}

\maketitle
\nomenclature{DT}{Digital Twin}
\nomenclature{DM}{Digital Model}
\nomenclature{DS}{Digital Shadow}
\nomenclature{P2D}{Data flow from Physical Twin to Digital Twin}
\nomenclature{D2P}{Data flow from Digital Twin to Physical Twin}
\nomenclature{D2D}{Data flow from Digital Twin to other digital systems}
\nomenclature{DEVOTION}{DEVelOping digital TwIn systems with autOmated model maNagement}
\nomenclature{MDE}{Model Driven Engineering}
\nomenclature{MBE}{Model Based Engineering}
\nomenclature{MBSE}{Model Based Systems Engineering}
\nomenclature{DSL}{Domain Specific Language}
\nomenclature{UML}{Unified Modelling Language}
\nomenclature{EMF}{Eclipse Modelling Framework}
\nomenclature{M2M}{Model to Model Transformation}
\nomenclature{M2T}{Model to Text Transformation}
\nomenclature{T2M}{Text to Model Transformation}
\nomenclature{DTME}{Digital Twin Management Environment}
\nomenclature{EOL}{Epsilon Object Language}
\nomenclature{EVL}{Epsilon Validation Language}
\nomenclature{ETL}{Epsilon Transformation Language}
\nomenclature{E/E Systems}{Electrical and Electronic Systems}
\nomenclature{SDTM}{Structured Digital Twin Metamodel}
\nomenclature{PWM}{Pulse Width Modulation}
\nomenclature{MC}{Model Checking}
\nomenclature{DTMC}{Discrete Time Markov Chain}
\printnomenclature[1in]

\section{Introduction}
\label{sec:introduction}

The concept of Digital Twin (DT) was originally proposed in the development of shuttles and has been widely embraced by researchers in various fields, such as aviation~\cite{li2021digital,wang2020application}, manufacturing ~\cite{kritzinger2018digital}, and urban constructions~\cite{sacks2020construction}. 
The escalating complexity of missions such as deep space exploration~\cite{yingzhuo2018china} and the development of reusable rockets~\cite{zhenxing2022rocket} demand a higher level of precision in modeling rockets and launch vehicles during both the ground-based system integration and autonomous control phases during flight, making DT a noteworthy prospective technology.
DTs offer distinct advantages in data monitoring of complex system, which can be harnessed for fault detection in spacecrafts~\cite{wenting2022preliminary}.
Furthermore, an increasing number of spacecraft designers are attempting to leverage machine learning and neural network algorithms, commonly employed in other engineering domains, to address challenges specific to rocketry ~\cite{huang2022comparative,xue2023research}.
Provided that these data-driven approaches require the management of extensive testing and flight data, DTs seem to be a viable solution for the above challenges.
However, due to the sheer scale and intricacy of spacecraft design, the current vague conceptualisation of DTs makes it challenging to develop a comprehensive DT framework for rocket systems, and the absence of suitable and specialised tools also impedes practical progress in this engineering domain.

The potential engineering benefits of Model Based Systems Engineering (MBSE) technology have gained widespread recognition among practitioners by promoting standardisation in complex systems engineering ~\cite{gregory2020long}.
As early as 2011, the National Aeronautics and Space Administration (NASA) initiated the application of MBSE in the aerospace domain and has been continuously advancing it.
For instance, the distinctive advantages of MBSE have been demonstrated in engineering experiments such as rocket engine testing and payload adapter design ~\cite{holladay2019mbse}.
Researchers from Airbus reviewed existing aerospace engineering development processes and proposed that MBSE could enhance interface consistency, communicability, clarity, and maintainability of aerospace electronic devices ~\cite{gregory2020long}.
In recent years, researchers at China Aerospace Science and Technology Corporation (CASC) have also recognised the advantages of MBSE in the field of space exploration ~\cite{wenyue2022application}. 
In the case of complex missions executed over multiple phases, such as lunar exploration missions, CASC's researchers applied MBSE technology to hierarchically analyse spacecraft mission processes, improving design integration and verification coordination~\cite{feng2022development}.

In this article, we perform a review on the concept of DTs and existing DT platforms and approaches to develop DTs. We then propose a DT maturity matrix, using which we could measure the maturity of DTs. We then propose DEVOTION (DEVelOping digital TwIn systems with autOmated model maNagement), which is a methodology which adopts model-based approach to develop DTs in any level of maturity. We then discuss our tool support, DTME (Digital Twin Management Environment), which is designed and developed specifically to support DEVOTION, with as many open source platforms as possible. We then report on a DT development case study, in which we develop a DT platform to assure the reliability of Electrical/Electronic Systems (E/E Systems) of space launch vehicles. 

The primary contributions of this paper are as follows:
\begin{itemize}
\item We clarify and consolidate the definition of DTs and introduce a DT maturity matrix, setting clear criteria when analysing the maturity level of DTs.
\item We propose a systematic, model based DT engineering methodology, DEVOTION, by following which, the developers may be able to develop a DT platform to any maturity level.
\item We discuss our tool support DTME, which is designed and developed to support the DEVOTION methodology, with a high degree of automation provided by model-based approaches..
\item We report our preliminary results on applying DEVOTION and using DTME to develop a DT platform for space launch vehicles, with a specific objective to analyse and assure the reliability of the Electrical/Electronic Systems (E/E Systems) for space launch vehicles.
\end{itemize}

\section{Background and Motivation}
\label{sec:preliminaries}

\subsection{Digital Twin}
The term "Digital Twin" was initially coined in 2003 when Grieves introduced the idea in his ``Product Lifecycle Management'' course ~\cite{grieves2014digital}. 
In his work, he described Digital Twins as comprising three fundamental elements: the physical product, its virtual counterpart, and the connections between them.
In 2012, the concept of DT was refined by the National Aeronautics and Space Administration (NASA). 
They defined Digital Twins as a \textit{multi-physics, multi-scale, probabilistic, ultra-fidelity simulation that reflects, in a timely manner, the state of a corresponding twin based on the historical data, real-time sensor data, and physical model}~\cite{glaessgen2012digital}.
According to Gabor et al.~\cite{gabor2016simulation}, Digital Twins are specialised simulations constructed using expert knowledge and real-world data gathered from existing systems. 
These simulations aim to achieve a more precise representation across various timeframes and spatial dimensions.
Tao and his team introduced a five-dimensional Digital Twin framework~\cite{tao2018digital} that includes the physical component, the virtual counterpart, the connections between them, data, and services.

As the concept of DT has gained popularity in academia and industry, systematic DT frameworks and DT maturity have been studied as important design starting points~\cite{sharma2022digital}.
In~\cite{zheng2019application}, an application framework of DT for product life cycle management in the manufacturing industry is proposed. The paper focuses on the technologies used in manufacturing information perception technology and presents a detailed discussion of the data storage, data processing and data mapping stages. The paper also touches upon the implementation process of parametric 3D modelling.
In~\cite{hu2023new}, the authors provide a DT maturity model for high-end equipment that combines qualitative and quantitative analysis. 
The proposed DT maturity model is applied to evaluate the DT maturity of three high-end equipment, including underground engineering equipment, a large wind turbine, and an industrial workshop. The evaluated maturity results could be used to increase the maturity of DTs by improving rubrics with high significance, low level and low improvement difficulty.
In~\cite{li2021digital}, the authors provide an in-depth discussion of DTs (Aero-DT) for the aerospace industry. The authors detail their state-of-the-art feature composition and the corresponding limitations to express the expectations of Aero-DT.
In~\cite{yang2021application}, the authors propose the concept of Spacecraft DT (SDT) and present a four-dimensional model conceptual structure adapted to the spatial distribution, where the key SDT issues are discussed, including data acquisition, system configuration, and data service modelling.
In~\cite{wenting2022preliminary}, the authors summarise research on digital twins in solid rocket motors, liquid rocket motors and rocket inertial device fault diagnosis, and propose a methodology and tool platform for driving rocket fault diagnosis with digital twins.
In~\cite{xiao2023multi}, using the aerospace product equipment workshop as a case study, the authors achieve multi-dimensional modeling of digital twin shop floor (DTS). Accurate DTS geometric modeling will be intelligently used for anomaly handling in manufacturing workshops.

\subsection{Model Based Systems Engineering (MBSE)}
The term MBSE has its origin in Model-Based Engineering (MBE), which emerged in 1970's in parallel with the evolution of Computer Aided Design (CAD) and Model Based Design (MBD) techniques. The main goal in MBE was to support the system develop process during the design, integration, validation, verification, testing documentation and maintenance stages~\cite{mittal2013model}. In systems engineering, the application of MBE principles is called as Model Based Systems Engineering (MBSE)~\cite{zeigler1993model,wymore2018model}. MBSE provides the required insight in the analysis and design phases, enhances better communications between different participants and enables effective management of the system complexity.

Model Driven Engineering (MDE) is a software development paradigm that uses models to support various stages of the development life cycle and can be seen as a subset of MBE~\cite{mittal2013model}. The core concepts of MDE include: a) the definition of Domain Specific Languages (DSL), in which the domain experts capture the concept of the system without needing to worry about underlying implementation technologies; and b) the development of model management operations, which can be executed on model in an automated manner, until the end software product is developed~\cite{wei2023automated,WEI2024112034}. In recent years, MDE had been applied in not only software engineering, but also other systems engineering involving also hardware ~\cite{jiang2021bridging,jiang2020re}. Hence, MDE can be also considered as a subset of MBSE.

\subsection{MBSE and Digital Twin}

\begin{figure}[h]%
\centering
\includegraphics[width=1\textwidth]{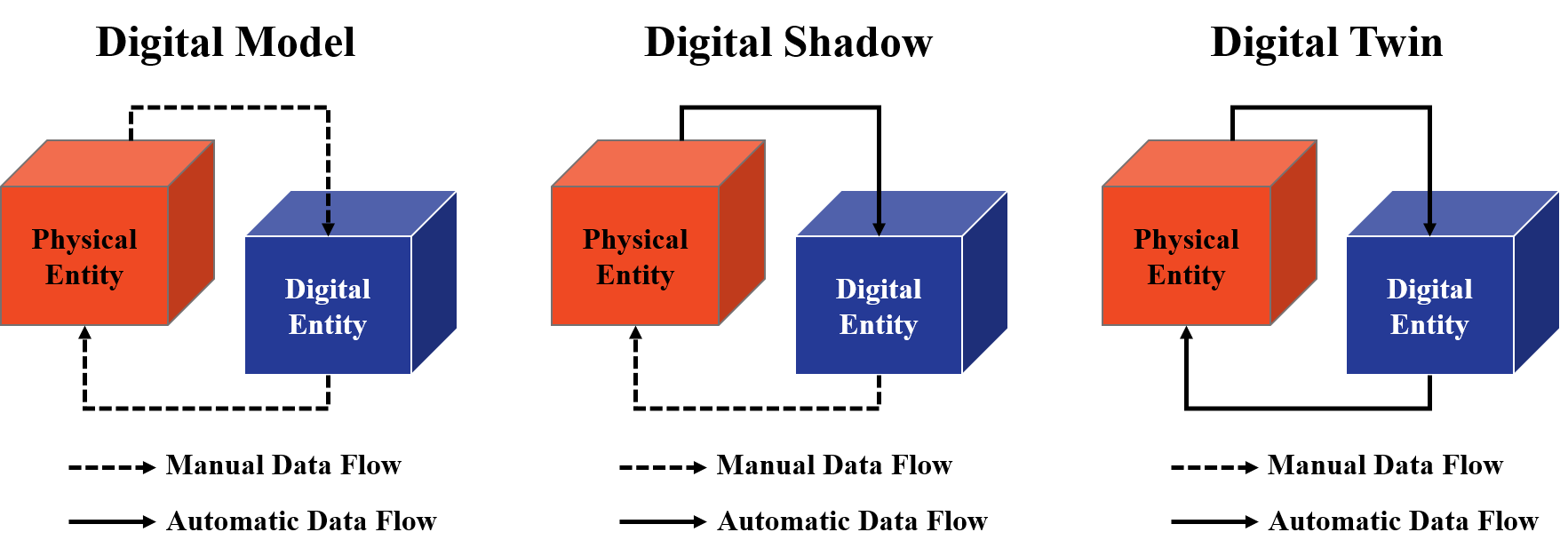}
\caption{The evolution of Digital Models into Digital Twins (Adapted from~\cite{kritzinger2018digital}).}
\label{fig1}
\end{figure}

MBSE is a promising technological route to achieving the goal of high-maturity digital twins, and there is a growing body of research that shares this view.
In~\cite{cencetti2013system}, The authors fully analyse the benefits of MBSE and in particular domain-specific language modelling in aerospace engineering and enumerated the benefits of MBSE.
In~\cite{schluse2017experimentable, zhang2022integrated}, the authors show the potential of MBSE for cross-discipline and cross-domain simulation, which provides strong support for achieving the goal of DTs.
In~\cite{bachelor2019model}, using the design of an ice protection system for a regional aircraft as an engineering story, the authors describe how the implementation of DTs can be developed and expanded by model-based design and standards such as FMI, which provide a reference for the fusion of heterogeneous models of complex systems.
In~\cite{schroeder2020methodology}, a systematic review of DT across industries is provided. The authors used MDE to capture the conceptual model used to construct a DT model. The authors also emphasise that a 3D geometric model is not necessarily part of the DT model and is merely a means of visualising the information contained within the DT model. The authors propose a DT reference architecture for the manufacturing industry that can be instantiated with AutomationML (AI based algorithm) to create the DT models that describe the PT.
In~\cite{bickford2020operationalizing}, using unmanned naval systems as an engineering story, the authors explore how the goals of DT development align with those of MBSE and how the MBSE process answers the questions needed to define DT. DT development is brought forward to an early stage in the system lifecycle by leveraging the work that has already been done in the system integration process.
In~\cite{madni2021digital}, the authors apply MBSE theory to the development of a DT system for UAVs that enhances system adaptive co-ordination, survival in the event of system failure and unexpected external disturbances, and other safety adaptive capabilities.
In~\cite{kritzinger2018digital}, the authors consider the digital replica of the physical world as Digital Models (DM). They also discuss the process of the DM into Digital Shadow (DS) and Digital Twin (DT), based on the extent of automation in data exchange between physical entities and their digital counterparts (as illustrated in Figure~\ref{fig1}).

\subsection{Digital Twin Maturity Matrix}
\label{sec:maturity}

\begin{table}[h]
  \begin{threeparttable}[h]
   \caption[\TeX{} engine features]{DT Maturity Matrix}
   \label{tab1}
   \centering
    \begin{tabular}{llll}
    \hline
    Maturity               & Data Flow\tnote{1}     & Automation & DM/DS/DT\tnote{2}  \\ \hline
    1. Descriptive         & None          & Minimal            & DM       \\
    2. Analytical          & None          & Limited         & DM       \\
    3. Operational         & P2D           & Moderate        & DS       \\
    4. Prescriptive        & P2D, D2P      & Substantial     & DT       \\
    5. Cognitive           & P2D, D2P      & High            & DT       \\
    6. Connected Cognitive & P2D, D2P, D2D & Complete        & DT       \\ \hline
    \end{tabular}
     \begin{tablenotes}
       \item [1] P2D: Physical to Digital, D2P: Digital to Physical, D2D: Digital to Digital.
       \item [2] DM: Digital Model, DS: Digital Shadow, DT: Digital Twin.
     \end{tablenotes}
  \end{threeparttable}
\end{table}

Based on various definitions of DT, together with the identification of the capabilities of the DT~\cite{autodesk, gartner, kpmg}, we propose a \textit{DT Maturity Matrix}, as shown in Table~\ref{tab1}. It provides a breakdown of the maturity levels, data flow directions (Physical to Digital - P2D, Digital to Physical - D2P, and Digital to Digital - D2), the degree of automation for the level, and the associated model classification, which includes Digital Model (DM), Digital Shadow (DS), or Digital Twin (DT). We now discuss the technological details needed for each level on the matrix.

\textbf{Level 1. Descriptive}. On this level, a descriptive model (i.e. a DM) of the physical entity (or system, we would use the term physical entity hereafter) should be in place, the descriptive model can be an information model that captures the concepts (their attributes, and their relationships with other concepts) within the physical entity. 
The level of abstraction for the descriptive model depends on the purpose of the DT, and should be adjustable for future needs.
On this level, no data from the physical entity is needed. 
Therefore, there is no data flow, and the level of automation is minimal. 
The model on this level can only be categorised as a DM.

\textbf{Level 2. Analytical}. On this level, the DM is enriched with other sources of data , including (but not limited to) laser scans, RGB images, thermal images, 3D models, simulation models, requirement models, etc. 
The data should be associated with the DM. 
In addition, analytical functions/algorithms shall be in place to analyse the DM and the results from the analyses shall be recorded in the DM, for humans to inspect the results (e.g. to determine if problems exist) and make decisions based on them. 
On this level, there is no automated data flow, the analytical algorithms promotes the degree of automation and the model is still considered to be a DM.

\textbf{Level 3. Operational}. On this level, means of synchronising sensor data with the DM should be in place, which enables the timely update of the DM. 
The frequency of the update depends on the application domain. 
On this level, the DM becomes a DS, the data captured for the physical entity is sent automatically to the digital entity, to update the conditions of the physical entity. 
However, for this to happen, there is also a need to automatically process the data and associate the data with the DM. 
Based on the historical data the DM can keep, it is also possible to establish a \textit{simulation} facility, which makes use of the historical data to perform simulations to assess impacts on designs, environmental conditions, etc. to help make decisions regarding the physical entity.

\textbf{Level 4. Prescriptive}. On this level, means of sending data to the physical entity should be in place, which allows the automated data transmission from the digital entity to the physical entity, and change the state/configuration of the physical entity (hence the word ``prescriptive''). 
This important automated data flow from the digital side to the physical side is what differentiates a DS and a DT. 
Therefore we recognise that prescriptive digital entity to be an important stage of the evolution from the DM to the DT.

\textbf{Level 5. Cognitive}. On this level, advanced algorithms such as data science and AI algorithms are applied in a much wider extent, from data analysis, decision making, performance optimisation and automated physical entity control. 
On this level, DT is fully functional and autonomous, automating every aspect of the life cycle of the physical entity.

\textbf{Level 6. Connected Cognitive}. On this level, the DT is able to connect to other digital entity (including other DTs) to obtain a wider range of data and extract insights on a larger scale, in order to make more informed decisions. 
In order to enable this, the DT should be able to exchange information with other digital entities in an automated manner, without any interoperability issues. 
In addition, the DT shall be capable of processing information on a much larger scale (requiring more computational power), in order to facilitate in time decision-making in an automated manner. 

\subsection{Definition of Digital Twin}

With our review on the literature on DT and our proposed DT Maturity Matrix, we consolidate on the definition of DT, in the context of this paper. 

\begin{enumerate}
    \item A digital twin is a digital model to capture the attributes, behaviours and states of a physical entity and its context (physical twin) at a certain level of abstraction of choice to fit the DT’s purpose. 
    \item Automated data flow from the physical twin to the digital twin is enabled to reflect the state of the physical twin in a timely manner. The digital twin should also \textit{prescribe} the state of the physical twin through automated data flow from the digital twin to the physical twin. 
    \item The digital twin may also be complemented by defined (manual and/or automated) processes, for analysis, simulation, reasoning, optimisation and decision making, to prescribe the physical twin in an automated manner. 
    \item (Optional) The digital twin may also connect to and extract information from other digital systems to facilitate its defined processes in a broader context.
\end{enumerate}

Following the definition of DT, we would also like to emphasis that when DT reached to maturity level 4 (Prescriptive) and above, it is no longer \textit{just a model}.
In addition to the digital model that holds the information about the Physical Twin (PT), the DT is also accompanied with communication channels that facilitate bi-directional data flows from the PT to the DT model. In addition, the processes applicable on the digital model, can exist independently (as processes models if a full Model-Based approach is taken), therefore not part of the DT. 
In this sense, the DT (or more accurately, the DT system) should compose: a) the digital model that holds information about the PT; b) bi-directional data synchronisation facility; and c) processes executable on the digital model functions on different DT maturity levels.

We would also like to remark on the misconception that the DT is simply a high-fidelity 3D model of the PT, from the above definition it is clear that the DT is more than that - a 3D model of the DT is only a means to curate the information in the DT, functions that facilitate decision-making (or performs the decision-making in an automated manner) may not need the visualisation of the 3D model of the PT at all.

\subsection{Challenges in Developing Digital Twin Systems}
\label{sec:challenges}
From the definition of DT and the DT maturity matrix, it can be perceived that developing a DT is an incremental practice. 
Developers for DT typically need to start with a Digital Model (DM), and advance the DM into a Digital Shadow (DS), and eventually into a Digital Twin (DT).
If the practice to evolve DM into a DT is taken using a purely manual approach, there would be some challenges at each of the stages.

For \textbf{Maturity Level 1}, the first challenge is to devise the model that holds the information and represents the state of the PT.
This typically related to identified challenges in terms of modelling and data fusion in previous studies~\cite{tao2018digital, chen2023advance}. 
The design of the DM shall consider factors such as a) \textit{extensibility}, in the sense that the DM shall be extensible to cater with the evolution of PT; b) \textit{modularity}, the DM shall be as modular as possible, as the DM may describe the PT on different levels of abstraction. In addition, different aspects of the PT shall be captured using different modules so that they can be exchanged independently; c) \textit{interoperability}, considering the eventual form of the DT would be a \textit{Connected and Cognitive} (maturity level 6), the information exchanged with other digital systems and other DTs shall be in a format that is accepted across DTs to avoid interoperability problems. d) \textit{traceability}, considering the heterogeneity nature of models within the DT (e.g. the digital model, processed data models such as geometric mesh, imagery, etc.), the DM shall be able to reference to heterogeneous models and act as the backbone for the DT. 

In addition to the challenge of creating the DM considering the above principles, it is also a labour intensive task to map the PT into the DM, provided the complex nature of PT (especially for systems built for space missions), problems such correctness (caused by human error) and prolonged modelling time shall be addressed before the DM can be used as the foundation for the DT.

For \textbf{Maturity Level 2}, the challenge is to fuse the sensor data with the DM before any analysis can be performed to yield results. 
As previously mentioned, if performed manually, this would be an error prone, labour intensive and time consuming process.

For \textbf{Maturity Level 3}, the challenge is to develop the P2D data links manually. 
As previously mentioned, the PT would be complex systems, and in order to get insights on different components of the PT, it is necessary to get hold of the data being passed among components in a timely manner (or even real-time) in order to analyse the status of the components. 
Without any automation, this task is also labour intensive, and time consuming.

For \textbf{Maturity level 4}, same challenge happens to develop the D2P data links manually.

For \textbf{Maturity level 5}, additional effort is needed to devise the communication protocols to the user groups of the DT system (i.e. humans and machines), and means to curate the information in an automated manner is needed, due to the same reason mentioned above.

For \textbf{Maturity level 6}, communication protocols to other digital systems and other DTs shall be devised in an automated manner, to cater with API changes and system evolutions.

\subsection{Discussion}
In this paper, we aim to address the above challenges by proposing DEVOTION, which is a development methodology adopting principles of MBSE and MDE. DEVOTION is then supported by our tool, DTME, which provides a high degree of automation, to improve the efficiency and consistency of the development of the DT platform.

\section{Proposed Approach}
\label{sec:access}
In this section, we discuss our engineering methodology -- DEVelOping digital TwIn systems with autOmated model maNagement (DEVOTION).
DEVOTION is a step-wise methodology that guides the forward development of DT (up to any maturity level).
To promote practicability of DEVOTION, and to make an attempt to address the challenges identified in Section~\ref{sec:challenges}, we adopt principles of Model Driven Engineering (MDE), to leverage benefits of automation brought by it.

\subsection{Model Driven Engineering}
For the development of DT, we adopt Model Driven Engineering (MDE), a subset of and a supporting paradigm for MBSE~\cite{mittal2013model}.
MDE is a contemporary software engineering paradigm, which was proposed to address the challenges posed by the increasing complexity of software systems~\cite{atkinson2003model}.
In MDE, \textit{model}s are first class artefacts, and through a number of automated model management operations, the software product can be developed with a high degree of automation. 
Therefore, the models \textit{drive} the engineering process of software systems.
There are two important aspects in MDE: \textit{Domain Specific Modelling} and \textit{Model Management}. 

\textit{Domain Specific Modelling} is a modelling methodology, in which domain experts of the system shall be in charge in capturing the concepts in their systems in the form of \textit{metamodel}s, or \textit{Domain Specific Languages} (DSL). 
The development of DSL allows domain experts to focus on the concepts of the system only, without needing to worry about low level implementation technologies . 
Once the DSL is devised, developers can create models that conform to the DSL to design their systems.
Existing languages and technologies that allow the creation of DSL are the Unified Modelling Language (UML)~\cite{uml}, the Systems Modelling Language (SysML)~\cite{friedenthal2014practical}, and the Ecore language provided by the Eclipse Modelling Framework (EMF)~\cite{steinberg2008emf}.

Once the models are in place, a variety of \textit{Model Management} operations can be performed on the models in an automated manner, until the eventual software system is built.
Such \textit{Model Management} operations include:
\begin{itemize}
    \item Model validation, to identify (and rectify) inconsistencies in the metamodel/model structures, as well as setting additional constraints for model elements (e.g. a particular lower-limit and upper-limit of values for certain sensors).
    \item Model-to-model (M2M) transformations, to transform models defined in a modelling technology into models defined in other technologies. M2M transformations are typically useful when systems evolve and need to adapt to new technologies.
    \item Model-to-text (M2T) transformation, to extract the information within the models and generate texts. M2T transformation is often used to generate source codes, documents, manuals, etc. 
    \item Text-to-model (T2M) transformation, to parse readily available texts (e.g. source codes) into models. This is typically used in reverse engineering, to extract a DSL from existing systems.
    \item Model comparison, to detect the difference among models before they are merged/integrated into a single model.
    \item Model merging, to integrate two or more models into a single model.
\end{itemize}

With the use of DSL and model management operations, system developers can derive their own DSLs for their systems, create models that conform to the DSLs, and then use a series of model management operations (often iteratively and in combination) to produce the eventual system. 
MDE has been proven to improve consistency and productivity significantly due to the automation provided by model management operations \cite{jaaksi2002developing, karna2009evaluating}. 

\begin{figure*}[th]
    \centering
    \includegraphics[width=1\linewidth]{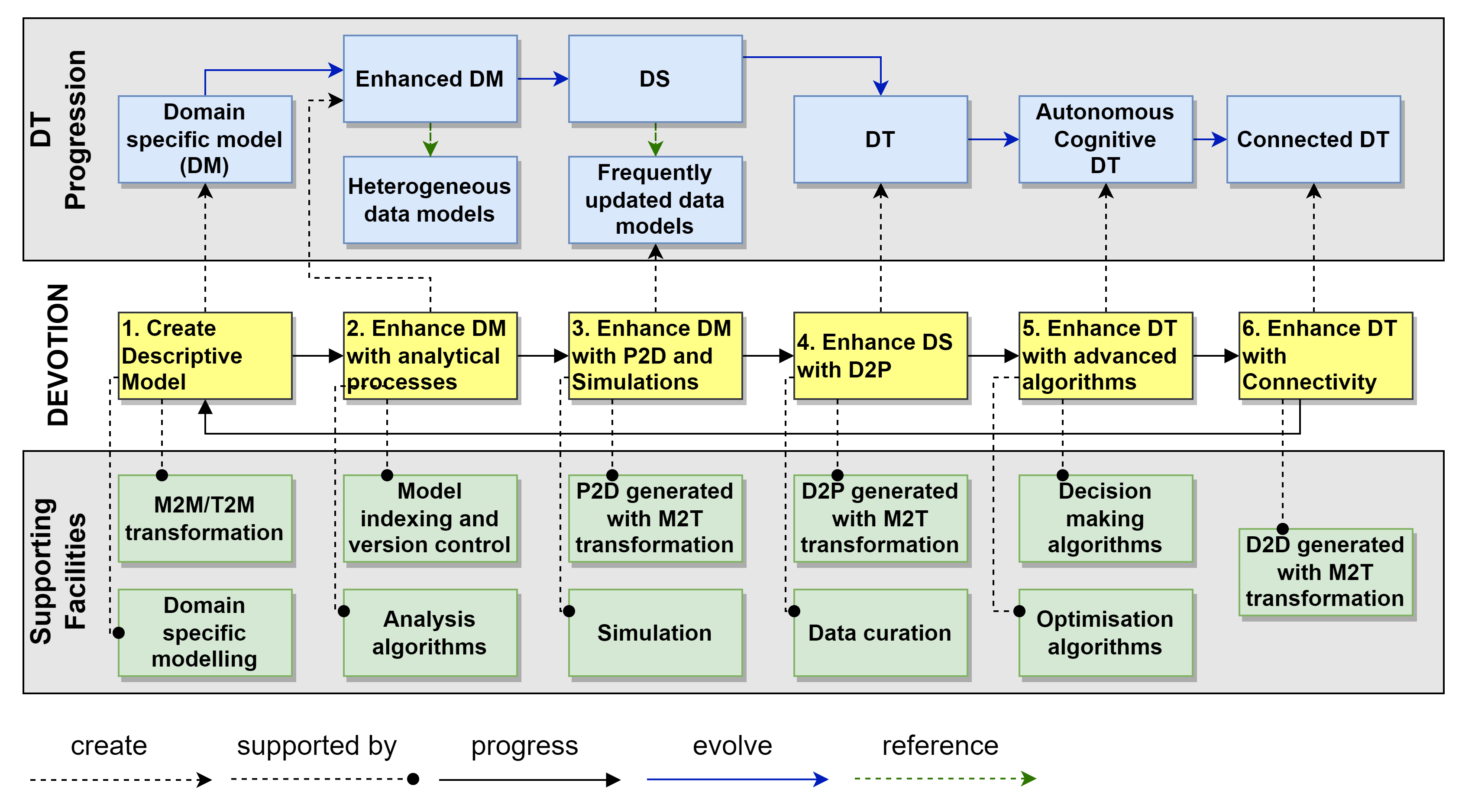}
	\caption{Stages and key models and processes for the DEVOTION methodology.} 
	\label{fig:devotion}
\end{figure*}

\subsection{Approach Discussion}
In this section, we discuss DEVOTION in detail, the individual steps involved in DEVOTION are shown in Figure~\ref{fig:devotion}. 
There are 6 steps in DEVOTION, which corresponds to the levels in the DT maturity matrix we presented in Section~\ref{sec:maturity}. 
At the same time, DEVOTION also suggest the types of models (upper swim lane), processes and facilities (lower swim lane) that should be in place for each step. 
We assume that model-based tool support is in place for DEVOTION, in order to address the challenges we identified in Section~\ref{sec:challenges}.

In \textbf{Step 1}, the descriptive model (i.e. the DM discussed in Section~\ref{sec:maturity}) of the PT should be created. 
For this purpose, it is important to obtain the DSL to \textit{describe} the PT through domain specific modelling. 
With the DSL in place, the next step is to see if there are readily available design models of the PT. 
If there are existing design models of the PT, M2M transformation can be used to transform the design models into the DM that conform to the DSL.
If there are no existing design models, and the source codes are available, it is possible to use T2M transformation to extract the DM that conform to the DSL.
The output of this step is the domain specific model (i.e. the DM) that conform to the DSL.
At the end of \textbf{Step 1}, the DT would have reached Maturity Level 1 as shown in Table~\ref{tab1}.

In \textbf{Step 2}, the DM can be enhanced with analytical algorithms and other data processing algorithms to perform analysis of the PT and infer their status.
This implies time-series data from the PT shall be collected (through sensors), processed and fused to the DM.
For this purpose, data processing algorithms shall be in place to convert the raw data in to processed data models.
Then, the processed data models shall be associated to the DM (i.e. the DM \textit{refer}s to the processed data models) to form an enhanced DM.
On the enhanced DM, analysis algorithms can be applied to infer the status of the PT and discover any other potential problems, such information is also stored in the enhanced DM. 
Upon checking the problems on the DM, developers can make decisions on how to alter and rectify the problems in the PT.
At this stage, model indexing and version control systems would help improve the scalability of the analysis algorithms, and at the same time provide facilities to query the model based on change and timestamp.
At the end of \textbf{Step 2}, the DT would have reached Maturity Level 2 as shown in Table~\ref{tab1}.

In \textbf{Step 3}, automated P2D data flow shall be in place to enable the automated timely (e.g. real time) synchronisation of PT data to the DM. 
This renders the DM into a DS as discussed in Section~\ref{sec:maturity} by our DT maturity matrix.
For this purpose, M2T transformation can be adopted to automatically generate the source code for communications for the P2D data flow. 
By enabling the P2D, PT data can be transmitted to the DS in a timely manner. 
In this step, it is also possible to deploy simulation algorithms to predict the behaviour of PT based on information in current/historical processed models. 
At the end of \textbf{Step 3}, the DT would have reached Maturity Level 1 as shown in Table~\ref{tab1}.

In \textbf{Step 4}, automated D2P data flow shall be in place to enable the automated data transmission from the DT to the PT.
This renders the DS into a PT as discussed in Section~\ref{sec:maturity}.
For this purpose, it is also possible to make use of M2T transformations to generate component-wise, D2P communications source code to send directives/instructions to the PT.
In addition, since the the state of the PT is synchronised to the DT through D2P and P2D data flows, it is also possible to curate the data, possible means of data curation could be a visualisation implemented in virtual environments (using 3D objects). 
At the end of \textbf{Step 4}, the DT would have reached Maturity Level 1 as shown in Table~\ref{tab1}.

In \textbf{Step 5}, the DT system is further enhanced with AI and other additional advanced cognitive algorithms. 
First, optimisation algorithms can be developed, which extracts the information from the DT (with its historical data), to determine the optimal level of performance, and generate instructions (to be sent to PT through D2P) on how optimal levels of performance can be achieved. 
In addition, advanced automated decision making algorithms can also be developed, which takes data in the DT as input, determine the configurations of the PT needed to achieve the optimal level of performance, and make decision to change the configuration of the PT (through D2P) in an automated manner.
At the end of \textbf{Step 5}, the DT would have reached Maturity Level 1 as shown in Table~\ref{tab1}.

In \textbf{Step 6}, the DT system can be enhanced to connect to other systems, including information systems, cyber-physical systems and DTs. 
In this sense, the DT system is able to obtain a wider breadth of information, making more accurate decisions.
For this purpose, M2T and M2M transformations are used to generate the source codes for D2D communications, and if the APIs are model-based, generate models for information exchange.
The enhancement of the DT system means advanced data handling capabilities, optimisation algorithms and decision making algorithms shall be developed to achieve decision making on information in a broader extent. 
At the end of \textbf{Step 6}, the DT would have reached Maturity Level 1 as shown in Table~\ref{tab1}.

By applying DEVOTION, it is possible to develop a fully working DT system from the start, up to any level of maturity, with a high degree of automation.
Should the PT evolve over time, with the model-based support, the DT system can be updated to cater with the changes of PT using model transformations in an automated manner.
The novelty of DEVOTION lies in the application of the MDE methodology, to ensure a high degree of automation, and at the same time promote consistency and interoperability.

\section{Tool Support}
\label{sec:toolsupport}
The DEVOTION methodology is backed by our model based tool support -- Digital Twin Management Environment (DTME). 
However, it is to be noted that DEVOTION can be supported by any model based tool, with the following key aspects mentioned in this section.
We invite the readers to note that the content covered in this section is for the case study (i.e. the twining of E/E systems of space launch vehicles), for a full twining system of space launch vehicles, a lot more tools and technologies shall be considered.

\subsection{Key Aspects}
\subsubsection{Domain Specific Language Development}
In DTME, the DSL is developed using Eclipse Modelling Framework (EMF).
EMF~\cite{steinberg2008emf} is the de-facto modelling framework in the context of MDE, such that a large number of tools in MDE is built atop EMF~\cite{barmpis2013hawk, kolovos2006eclipse, viyovic2014sirius}.
Therefore, a large number of open source model management tools are readily available for models created using EMF.
EMF provides a modelling language named Ecore, which allows the creation of DSLs for domain experts. 

Built within DTME, for the purpose of the discussion over the case study, a DSL named Structured Digital Twin Metamodel (SDTM) is defined using EMF's Ecore. 
SDTM is a highly modular and extensible DSL, and is used to create the digital twin model for this work.

\subsubsection{Model Management}
In DTME, model management operations are created and executed against the DT model using the Eclipse Epsilon platform~\cite{kolovos2006eclipse}.
Epsilon is an integrated model management platform which supports the management of models defined in arbitrary modelling technologies in an automated manner.
Epsilon provides a family of interoperable model management languages, such as the Epsilon Object Language (EOL) for model query and creation, the Epsilon Validation Language (EVL) for model validation, and the Epsilon Transformation Language (ETL) for model transformation.
In addition, Epsilon provides a set of model drivers in Epsilon Model Connectivity (EMC), which allows the access of models defined in different modelling technologies (e.g. EMF, Excel Spreadsheet, plain XML and Simulink). 
In addition, EMC is also extensible, and model \textit{driver}s can be developed to access metamodels/models defined in any modelling technology.

DTME uses Epsilon's EMC to access models defined using different modelling technologies (in our case study, EMF, Simulink and Excel models), and makes use of the Epsilon languages to perform model management operations used within the DEVOTION process.

\subsubsection{Modelling Environment for DSL}
With the DSL in place, a modelling environment is needed to create and manage models that conform to the DSL in an efficient manner.
For DTME, we make use of the Eclipse Sirius~\cite{viyovic2014sirius} platform.
Sirius enables the users to create a graphical modelling workbench by leveraging EMF. 
By defining \textit{ViewPoints} in Sirius, users are able to create complex graphical modelling editors with complex functionalities.

In DTME, we create a graphical modelling framework, which facilitate the creation of digital models that conform to SDTM, as well as the M2M transformation to transform system design models into instance models that conform to SDTM.

\subsubsection{Legacy Models and Simulation}
In some cases, it is not necessary to build the DT from scratch, some legacy models can be used. 
In our case study, we show that some existing designs of the E/E systems are already in place in Matlab/Simulink.
Simulink provides a graphical block-based modeling framework that facilitates the creation, simulation, and examination of systems.
In our work, we show: a) how existing/legacy models can be used to create models that conform to our DSL; b) how heterogeneous models (i.e. models defined in different technologies such as Simulink) can be linked together with the mechanisms provided by DTME; and c) how other tools such as Simulink can be used to perform simulations of the system.

\subsubsection{Model Storage and Version Control}
As the level of detail increases for the DT system, together with the increasing amount of data (both historical and current), the DT model will inevitably become very large. 
Consequently, file-based models are no longer suitable for tools to perform queries, analyses and transformations on them in a scalable manner. 
On the other hand, data science and AI algorithms typically need (a large amount of) history data to extract insights to inform decision-making, therefore version control systems are needed to get the state/conditions of the PT (as a whole) at points in time.

To improve scalability and version control, in this work we adopt Eclipse Hawk~\cite{barmpis2013hawk}, which is a scalable model indexing framework, that monitors version-controlled (using Git) model repositories and indexes all the contents in the repositories into a database back-end. By doing so, queries on models are executed against the back-end, eliminating the need to load models into memory and perform queries on in-memory models. 
Hawk also provides timestamp based queries, so that the data science and AI algorithms may query Hawk to get the data needed in time series for mining and training purposes.

\begin{figure*}[h!]
    \centering
    \includegraphics[width=1\linewidth]{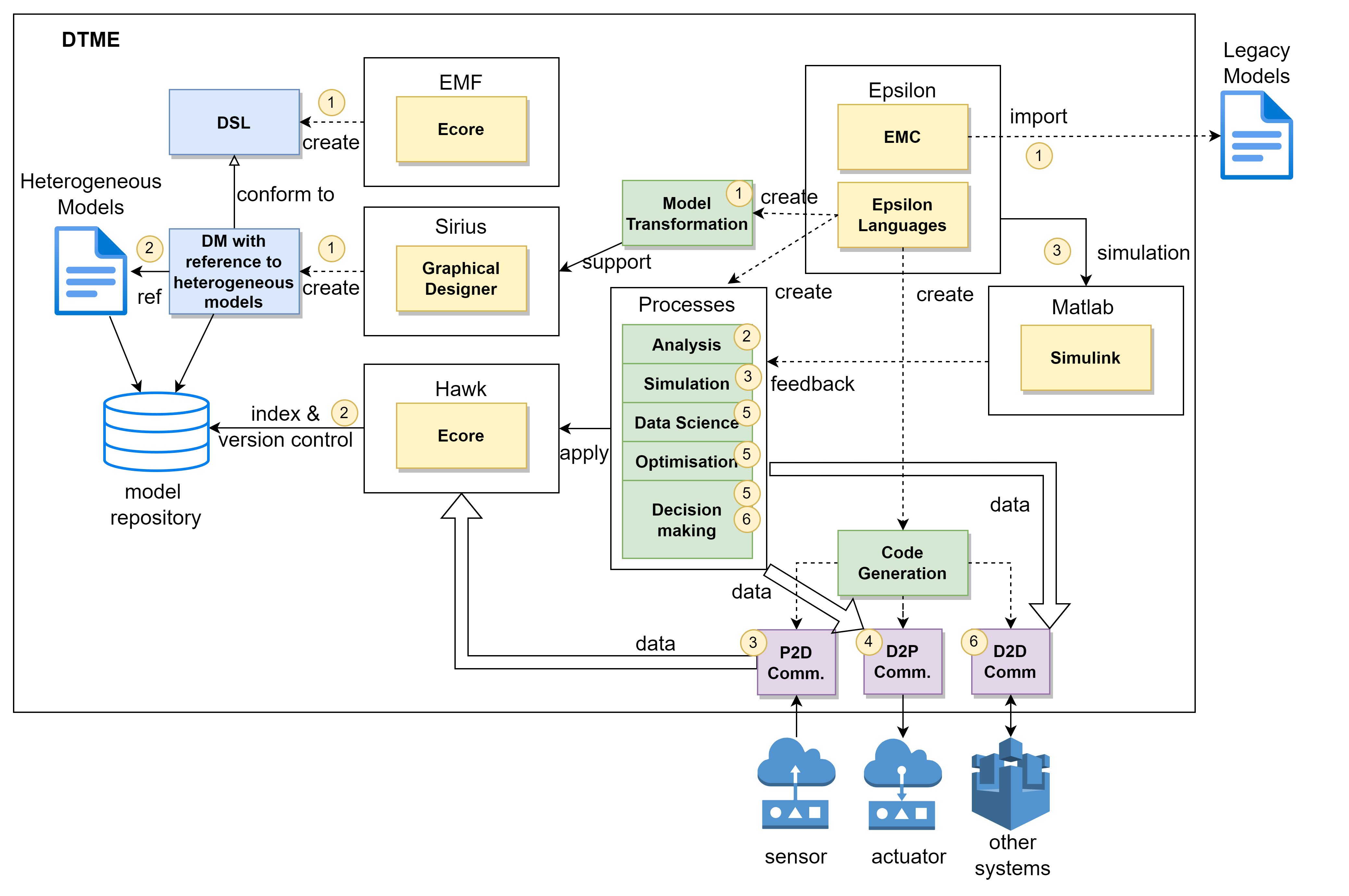}
	\caption{Architecture of the Digital Twin Management Environment (DTME).} 
	\label{fig:dtme}
\end{figure*}

\subsection{Tool Working Process}
We now discuss the working process of DTME in relation to the DEVOTION process.
Figure~\ref{fig:dtme} shows the working process of DTME in which circles with numbers in them correspond to the DEVOTION steps.

In DTME, we make use of EMF for domain specific modelling, Eclipse Sirius to create graphical model editor, Eclipse Hawk for model indexing and version control, Eclipse Epsilon for model management and Matlab Simulink for simulation, as shown in Figure~\ref{fig:dtme}. 

In \textbf{Step 1}, the DSL named SDTM is created using EMF's Ecore modelling language. 
With the SDTM in place, we are able to create a graphical modelling environment use Eclipse Sirius (as shown in Figure~\ref{fig:dtme_demo}), which provides ease to create instance models of SDTM, and at the same time, we are able to incorporate additional functionalities in it. 
For example, it is possible to \textit{import} legacy models, by using M2M transformations to covert the legacy models into instances of SDTM. In our tool, we provide a M2M transformation to transform Matlab Simulink system design models in to instance models of SDTM.

In \textbf{Step 2}, models that contribute to the digital replication of the PT shall be fused together. Such models are heterogeneous in nature, and may include: a) 3D geometric models that replicate the PT, which may be obtained by laser scanner and then converted to geometric mesh models; b) sensor data such as imagery and readings of E/E systems, which can be to reflect and infer the current state of the PT; c) engineering models such as system requirement models and system design, the change of which would inform that the DT needs to be updated; d) simulation models, which may be executed with the current configuration of the system to predict the condition of the system.
To fuse the data in the above models, traceability links needs to be established from the SDTM model to the heterogeneous models. 
In this work, the traceability is established through a built-in mechanism of SDTM, and the retrieval of the information through the traceability shall be realised through model management (i.e. model queries) written in Eclipse Epsilon.

The traceability that exist between the SDTM and heterogeneous models logically forms a graph, which may be difficult to maintain as the model grows large as discussed in previous sections. 
In DTME, we put all of the models in a version controlled (using Git) repository, which is monitored by Eclipse Hawk. 
Eclipse Hawk is able to detect the changes in the repository and index the changes into a back-end (currently using a graph database). In this way, queries on the DTME model (and its referenced heterogeneous models) would be executed against Eclipse Hawk, which provides scalable and efficient query results.

In \textbf{Step 3}, with the enhanced DM in place, the next step is to establish the P2D data flow. In our work, we fully exploit the benefits brought by MDE and use M2T generation to generate the source code for the communication with the PT. To do this, the developer simply needs to mark the model elements that should receive data, and corresponding communication classes (written in Java) will be generated.

With timely updated data through P2D, it is also possible to invoke simulation with the current configuration of the system and determine the system state in the future. In our work, we incorporate within DTME the ability to invoke the simulation function within Simulink, to illustrate how this could be achieved.

In \textbf{Step 4}, D2P APIs are automatically generated by M2T transformations, with the similar approach in \textbf{Step 3}.

In \textbf{Step 5}, value-added services such as data science, optimisation and decision making algorithms can be developed. We do not put restrictions on how the above algorithms are implemented -- Eclipse Hawk provides APIs implemented with Apache Thrift, it implies that the above algorithms written in any language may use the data from Hawk (as a result of executing queries to Hawk). 

In \textbf{Step 6}, D2D APIs can be generated again using M2T transformations so that the DT platform is able to communicate with other systems, but it is to be noted that this would require further investigations as we have not seen any DT implementation that have reached this level of maturity.

In this paper, we are not ambitious and aim to cover all DT maturity levels with our tool support. 
Instead, we illustrates the feasibility of our methodology and tool support. 
We would focus on the discussion on DEVOTION steps 1-5 for in our case study in the following section.

\section{Case Study}
\label{sec:case}
In this section, we discuss a case study on developing a DT for an E/E system of a space launch vehicle, to illustrate the DEVOTION methodology and DTME.
For illustrative purposes, we discuss the DT development, starting with a DM (DEVOTION step 1) up until the DM evolves into a DT (DEVOTION step 4 and 5).
In this example, we focus on a direct current motor speed control system based on (Pulse Width Modulation) PWM technology in an electric power transmission system.
Electrical system engineers developed a model using Matlab/Simulink and performed calculations according to actual engineering requirements. 
The system's hardware includes a PWM pulse generator, an H-bridge driving main circuit, a DC motor, and a feedback filter. 
Additionally, the closed-loop controllers all employ PID algorithms. 
In traditional simulations, the effectiveness of the power system can be verified by adjusting control design and controller parameter design. 
During actual testing, the entire system is treated as a whole for testing to compare the final results with the simulation results.

\subsection{DEVOTION Step 1: Creation of SDTM}
\label{sec:sdtm}
In DEVOTION \textbf{Step 1}, we create the Structured Digital Twin Metamodel (SDTM) using EMF, which is the DSL that drives the development of the DT in our case study.
SDTM is an extensible DSL that allows the creation of DMs to describe systems in space launch vehicles on different abstract levels. 
The focus of this work is on the reliability of E/E Systems of launch vehicles, therefore we will focus on parts of SDTM related to E/E Systems.
When designing the SDTM, we adhered to the following principles:
\begin{itemize}
    \item \textbf{Modularity}. SDTM shall be modular to capture the different aspects of a digital twin. In general terms, the SDTM shall cover the terminologies in the chosen DT domain, the artifacts involved in the DT. In domain specific terms, the SDTM should organise the DT model in separate packages, so that they can exist independently, for reuse and interchange.
    \item \textbf{Extensibility}. Components of SDTM are extensible, SDTM contains a Base package, which allows the users to extend SDTM and adapt SDTM to their own needs with minimal efforts.
    \item \textbf{Traceability}. SDTM is designed to enable the traceability from SDTM models to external and heterogeneous models (models defined in different technologies), so that a SDTM model can act as a federation model to integrate system information (data fusion).
\end{itemize}

Consequently, SDTM contains five packages: the \textit{Base} package to capture common attributes of DT model elements; the \textit{Terminology} package to capture the terminologies used in the context of DT; the \textit{Artifact} package to capture the kinds of artifacts involved in the context of DT; the \textit{Component} package to capture the components of E/E systems for space launch vehicles; and the overall \textit{DigitalTwin} package to integrate the above packages into a DT.

\begin{figure}[th]
    \centering
    \includegraphics[width=1\linewidth]{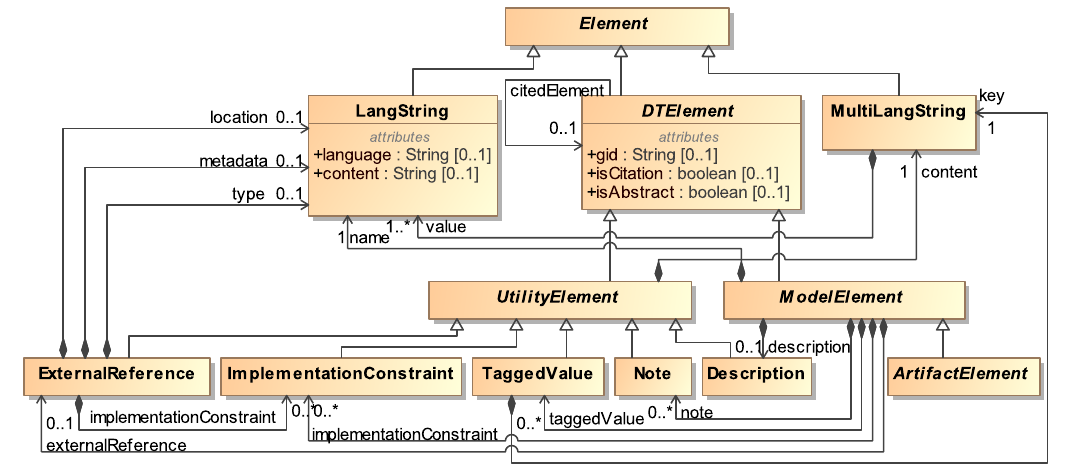}
	\caption{The Base package of the Structured Digital Twin Metamodel (SDTM).} 
	\label{fig:base}
\end{figure}

\subsubsection{The Base package}
The \textit{Base} package of SDTM is shown in Figure~\ref{fig:base}. 
The base package is relatively complex since facilities are designed to promote modularity, extendability and traceability.
The core SDTM element is the \textit{ModelElement}, where it has a \textit{name} (of type \textit{LangString}, which allows the users to specify a string as well as the language used), and a number of \textit{UtilityElement}s, which are:
\begin{itemize}
    \item \textit{Description}, to provide descriptive information for the \textit{ModelElement};
    \item \textit{TaggedValue}, which allows the users to associate key-value pairs to \textit{ModelElement};
    \item \textit{Note}, which allows the users to attach a note to the \textit{ModelElement};
    \item \textit{ImplementationConstraint}, which allows the users to attach a series of constraints (including machine-executable constraints) to the \textit{ModelElement};
    \item \textit{ExternalReference}, which allows the \textit{ModelElement} to refer to information that exists outside the SDTM model that contains the \textit{ModelElement} (e.g. to refer to a reliability model). In the \textit{ExternalReference}, the users are able to specify: the \textit{location} of the external model, the \textit{type} of the external model, the \textit{metadata} that describes the external model (if it exists), and finally the (machine-executable) \textit{implementationConstraint} which, when executed, are able to extract information from the external model. 
\end{itemize}
A \textit{ModelElement} is also able to ``cite'' another \textit{ModelElement}, in the sense elements inside a SDTM model may have traceability to elements that may be organised in another package.
With the above \textit{UtilityElement}s, a \textit{ModelElement} is able to provide multi-language support, as well as traceability to external models.

\begin{figure}[h]
    \centering
    \includegraphics[width=1\linewidth]{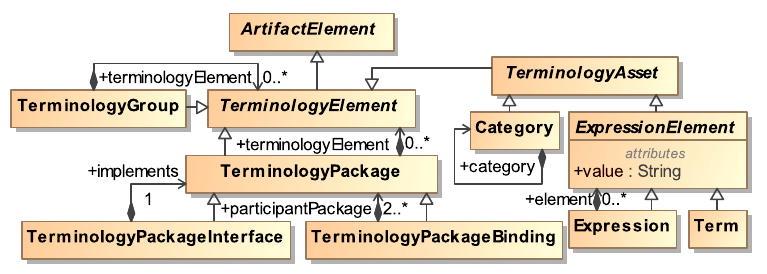}
	\caption{The Terminology package of the Structured Digital Twin Metamodel (SDTM).} 
	\label{fig:terminology}
\end{figure}

\subsubsection{The Terminology package}
The \textit{Terminology} package is shown in Figure~\ref{fig:terminology}.
The \textit{Terminology} package extends the \textit{Base} package (as its root elements extends the \textit{ArtifactElement} in the \textit{Base} package).
The main elements of the \textit{Terminology} package are:
\begin{itemize}
    \item \textit{Term}, to capture a terminology used within the DT context.
    \item \textit{Expression}, to capture an expression with specific semantics used within the DT context.
    \item \textit{Category}, to group \textit{Expression}s and \textit{Term}s into categories.
\end{itemize}
All terminology related elements are organised in \textit{TerminologyPackage}s, to promote modularity. 
A \textit{TerminologyPackage} may have a number of \textit{TerminologyPackageInterface}s and \textit{TerminolgyPackageBinding}s, to promote the reuse, interchange and integration of \textit{TerminologyPackage}s.

\begin{figure}[th]
    \centering
    \includegraphics[width=1\linewidth]{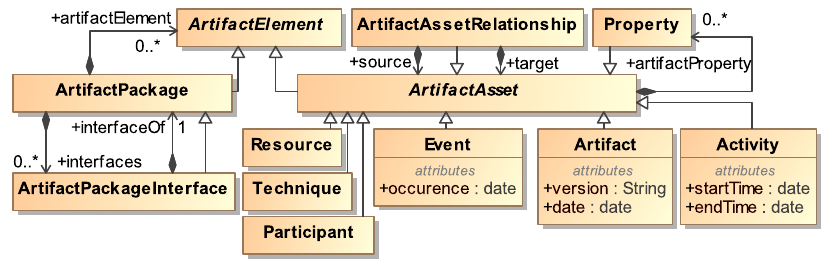}
	\caption{The Artifact package of the Structured Digital Twin Metamodel (SDTM).} 
	\label{fig:artifact}
\end{figure}

\subsubsection{The Artifact package}
The \textit{Artifact} package is shown in Figure~\ref{fig:terminology}.
The \textit{Artifact} package extends the \textit{Base} package (as its root elements extends the \textit{ArtifactElement} in the \textit{Base} package).
The main elements of the \textit{Artifact} package are:
\begin{itemize}
    \item \textit{Artifact}, to capture an artefact in the context of the DT, its version and date of creation.
    \item \textit{Activity}, to capture an activity that happened within the DT, with a start time and end time.
    \item \textit{Event}, to capture an event that happened in the operation of the DT.
    \item \textit{Resource}, to capture the available resources within the DT.
\end{itemize}
All artefact related elements are organised in \textit{ArtifactPackage}s, to promote modularity. 
A \textit{ArtifactPackage} may have a number of \textit{ArtifactPackageInterface}s and \textit{ArtifactPackageBinding}s, to promote the reuse, interchange and integration of \textit{ArtifactPackage}s.

\begin{landscape}
\begin{figure}
    \centering
    \includegraphics[width=1\linewidth]{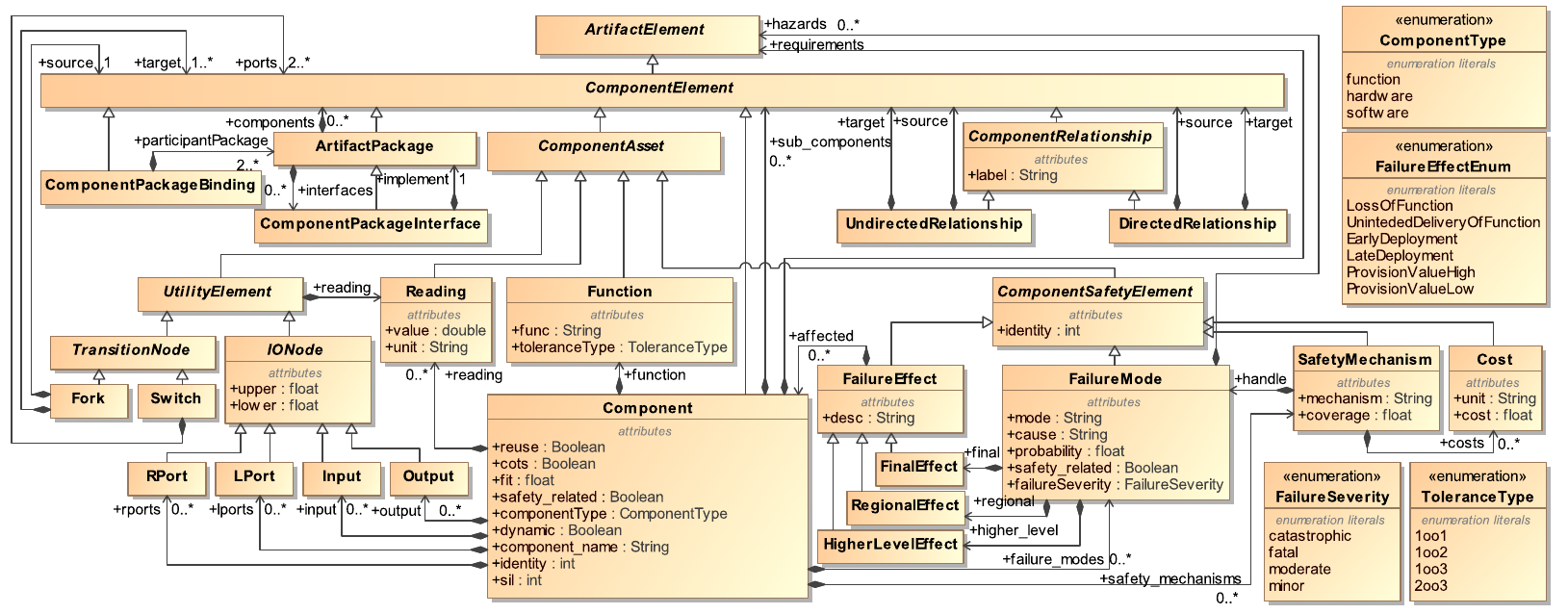}
	\caption{The Component package of the Structured Digital Twin Metamodel (SDTM).} 
	\label{fig:component}
\end{figure}
\end{landscape}

\subsubsection{Component package}
The \textit{Component} package is shown in Figure~\ref{fig:component}.
Again, the \textit{Component} package extends the base component.
Element \textit{ComponentElement} is the abstract for all architecture-related elements, and \textit{ComponentElement}s are organised in \textit{ComponentPackage}s, which in turn may have a number of \textit{ComponentPackageInterface}s.
The \textit{ComponentElement} allows the users to model:
\begin{itemize}
    \item \textit{Component}, which represents an atomic component in the user's systems. It may be \textit{reuse}d, or it may be \textit{COTS} (Commercial Of The Shelf) component. It may have a \textit{FIT} (Failure-In-Time, $10^{-9}$ failures/hour). It also has a \textit{safety integrity level}~\cite{jiang2018bluevisor}, which implies different levels of rigours for different application domains. Then the component may also have a \textit{ComponentType}, which may be system, hardware or software.
    \item \textit{ComponentRelationship}, which connects two \textit{Component}s.
    \item \textit{Function}, which may have a \textit{tolerance type}: 1oo1 (1 out of 1), 1oo2, 1oo3 or 2oo3.
    \item \textit{Input and Output}, to capture the inputs and outputs of \textit{Components}.
    \item \textit{FailureMode}, to capture the failure modes of a \textit{Component}.
    \item \textit{FailureEffect}, which allows the users to capture the effect of the failure. \textit{FailureEffect} may be used to refer to another \textit{Component} which would be affected. \item \textit{Component} by using the ``cite'' reference as described in the base component. As far as automated FMEA is concerned, computing a transitive closure of the effect of the \textit{FailureMode} would determine if a \textit{FailureMode} is safety-related.
    \item \textit{SafetyMechanism}, to capture the safety mechanism that can be deployed on a \textit{Component} to achieve diagnostic coverage.
\end{itemize}

\begin{figure}[h]
    \centering
    \includegraphics[width=1\linewidth]{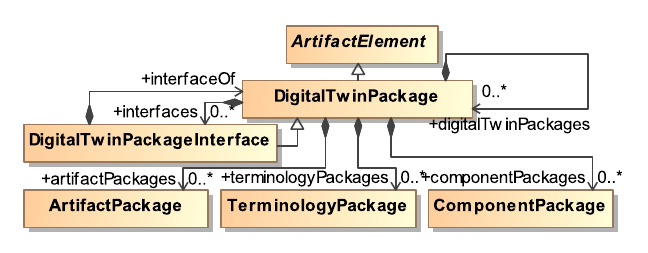}
	\caption{The Digital Twin package of the Structured Digital Twin Metamodel (SDTM).} 
	\label{fig:dtpackage}
\end{figure}

The \textit{Component} package is modelled specifically for the purpose of reliability and safety of the E/E system of space launch vehicles.
Users of SDTM are free to extend it to model other aspects of their systems, in this paper we will only focus on the modelling of the E/E systems of space launch vehicles.

\subsubsection{Digital twin package}
Putting the above packages together, the \textit{Digital Twin} package is shown in Figure~\ref{fig:dtpackage}.
In the current form of SDTM, a \textit{DigitalTwinPackage} may have a number of \textit{ArtifactPackage}s, \textit{TerminologyPackage}s and \textit{ComponentPackage}s. 
A \textit{DigitalTwinPackage} may also have a number of \textit{DigitalTwinPackageInterface}s for model exchange.

\begin{figure*}[h]
    \centering
    \includegraphics[width=1\linewidth]{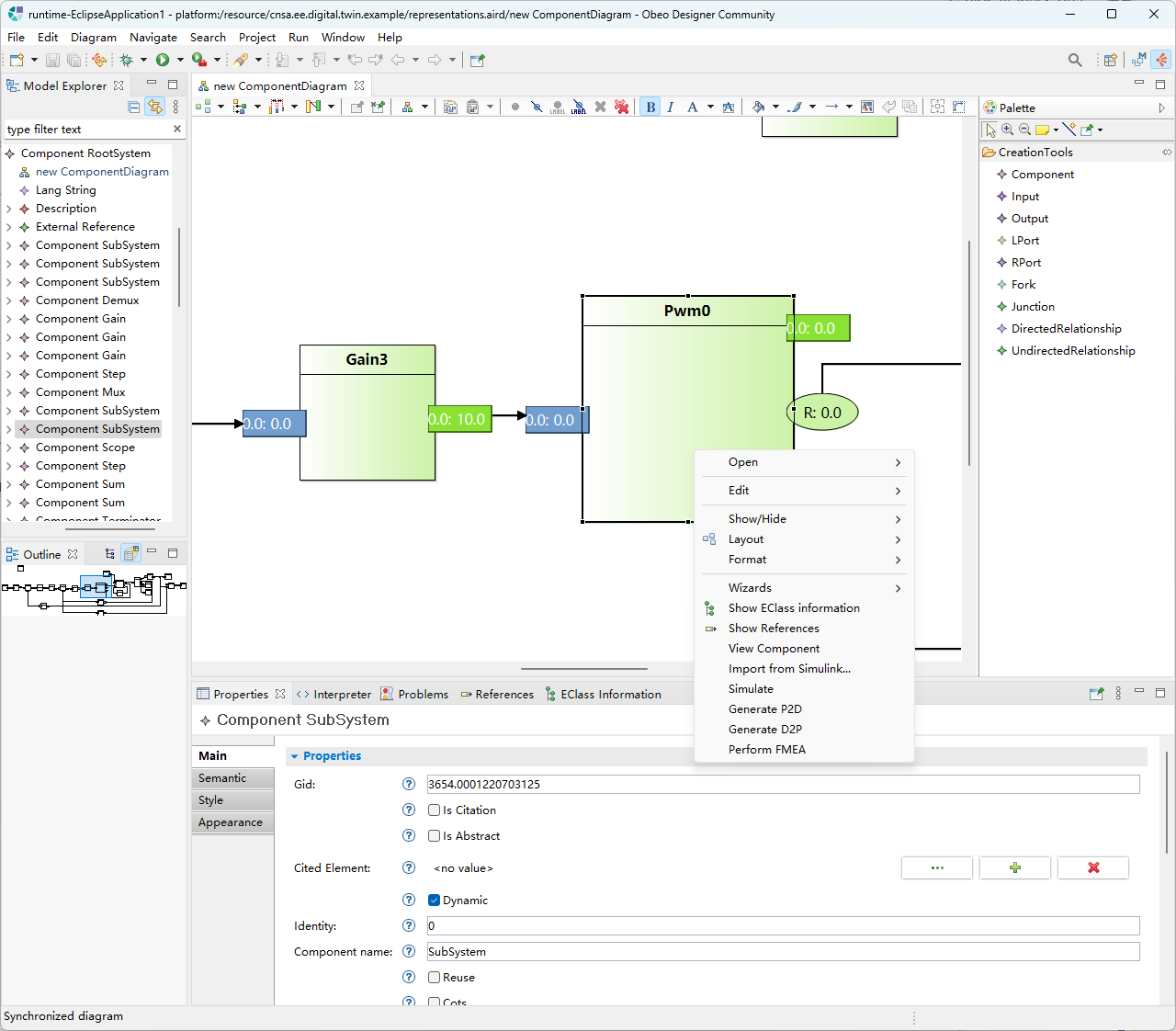}
	\caption{Graphical editor of SDTM integrated into DTME.} 
	\label{fig:dtme_demo}
\end{figure*}

\subsubsection{Graphical editor for DTME}
\label{sec:tool_SDTM}
With DTME defined using EMF, we developed a graphical modelling tool for SDTM using Eclipse Sirius~\cite{viyovic2014sirius}.
The tool support provides hierarchical graphical editors, which allows the users to create different packages of SDTM discussed in previous sections. 
Figure~\ref{fig:dtme_demo} illustrates the graphical editor for a \textit{Component}, in which we use block-based representation, and model specifically the I/O nodes of the \textit{Components}. 
It is to be noted that the shapes in the editors are customisable through Sirius, which is not the focus of this work.

\begin{landscape}
 \begin{figure}
  \centering
  \includegraphics[width=1\linewidth]{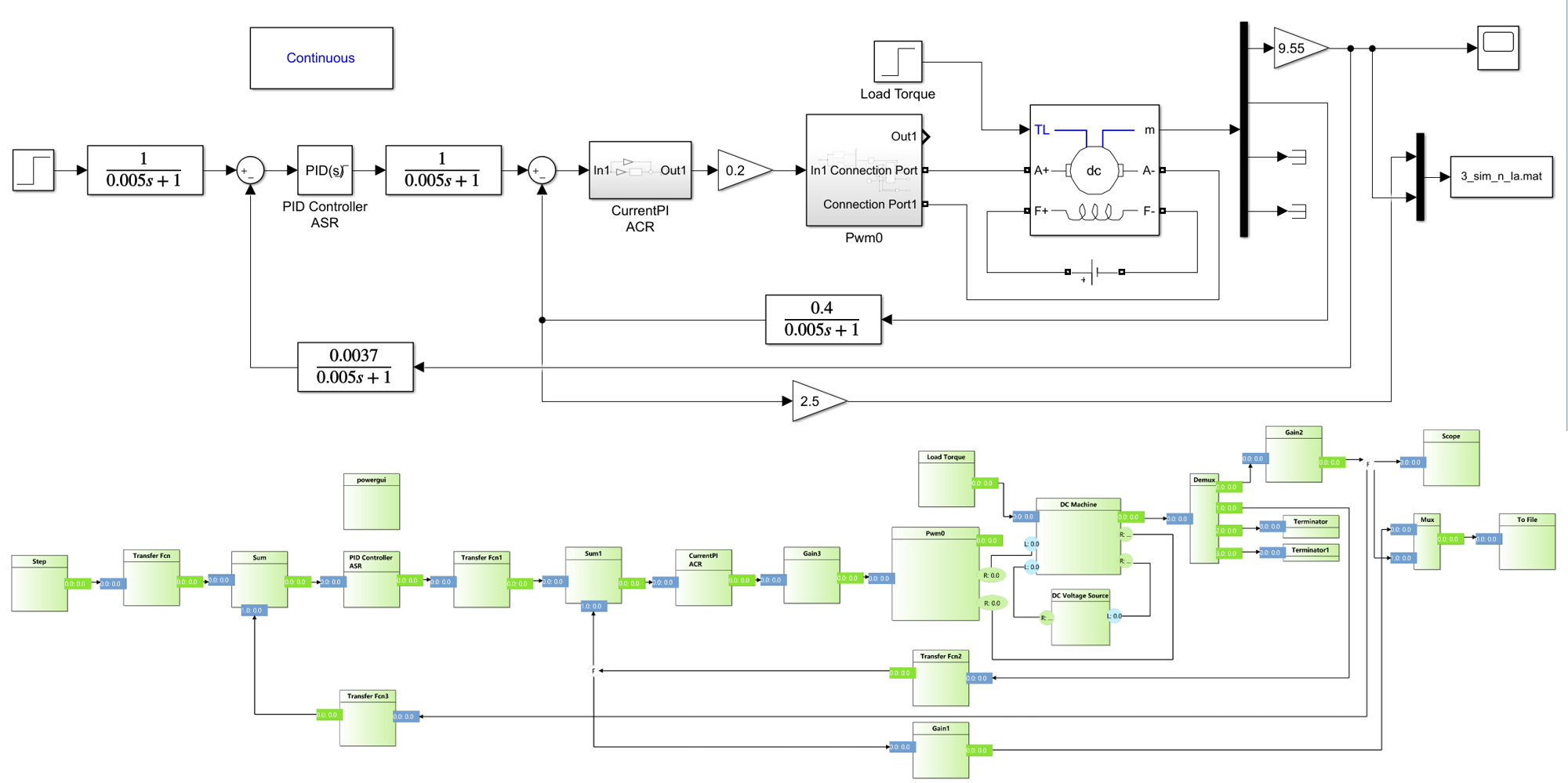}
  \caption{The one to one mapping from Simulink model to SDTM model.}
  \label{fig:m2m}
 \end{figure}
\end{landscape}

With SDTM and its graphical editor in place, our next objective is to create the DM that describes the motor speed control system.
Since the control system is a already designed in Matlab Simulink, to fully exploit the benefits of MDE, we developed a M2M transformation algorithm, which is able to transform Simulink models into DTME models.
The Simulink model, as well as the transformed DTME model are shown side by side in Figure~\ref{fig:m2m}.
Similar to Simulink, SDTM enables the modelling of nested systems (e.g. system of systems). 

\begin{figure*}[h]
    \centering
    \includegraphics[width=1\linewidth]{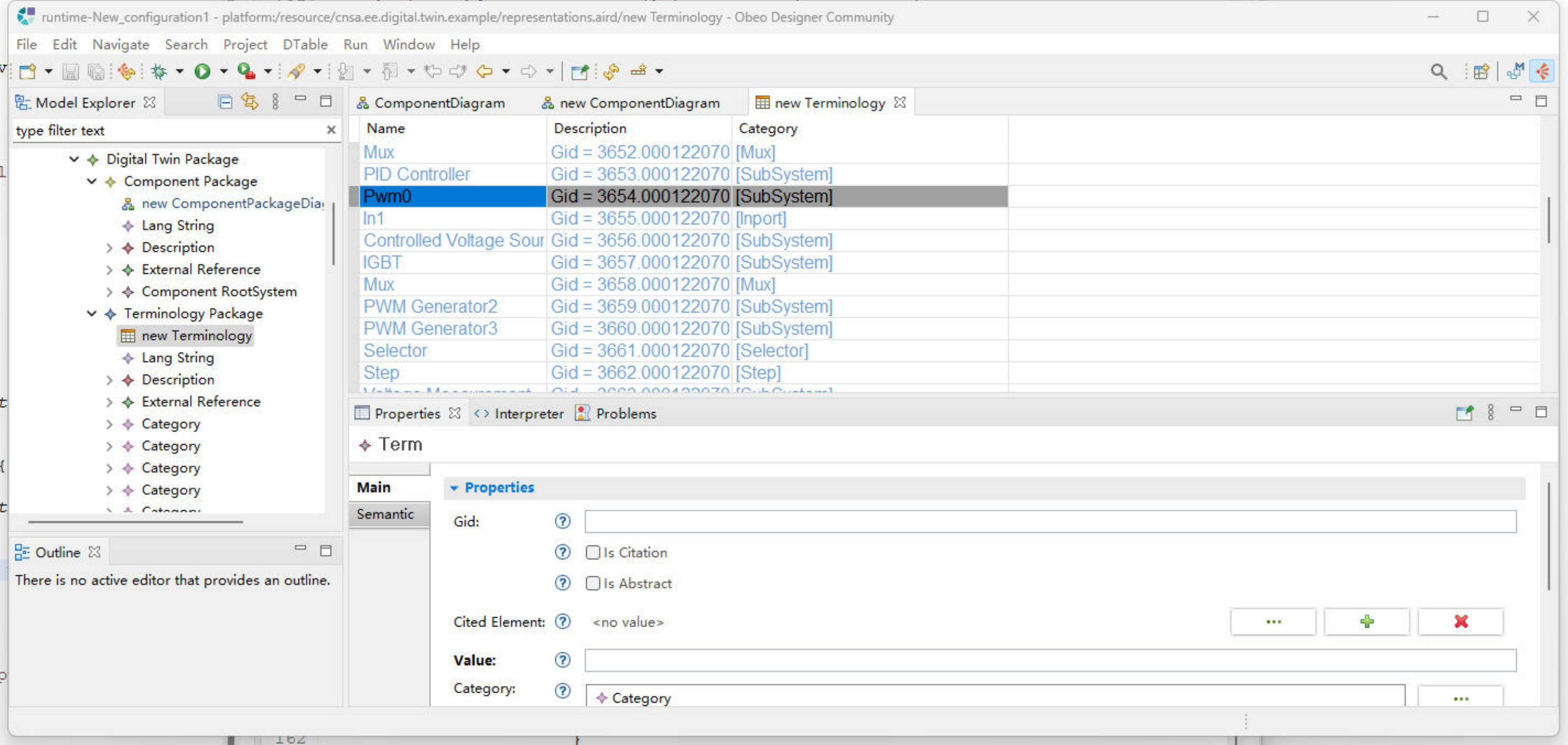}
	\caption{Component types as \textit{Terms} stored in a \textit{TerminologyPackage}.} 
	\label{fig:term}
\end{figure*}

It is to be noted that in the \textit{Component} package of DTME, we do not model \textit{type}s and hard code them in DTME, this is due to the fact DTME is designed to be applicable on models in different levels of abstraction. 
Instead, we store the type information of \textit{Component}s in \textit{Terminology} packages (each type is a \textit{Term} in the \textit{Terminology} package), as shown in Figure~\ref{fig:term}.
Such Terminology packages can also be exchanged amongst different DT platforms for consistency and interoperability.

\subsubsection{Discussion}
Up to this point, we reached Maturity Level 1 of our DT by having a DM, which is created using our defined DSL (DTME).
To any DT platform, it is crucial to get the DSL correct, as it will be the backbone to hold the information of the DT, and also the instance models of the DSL will be used as input for a series of model management operations, which can be used to perform analysis and generate source codes in an automated manner, boosting the efficiency in DT development.

\begin{figure*}[h]
    \centering
    \includegraphics[width=1\linewidth]{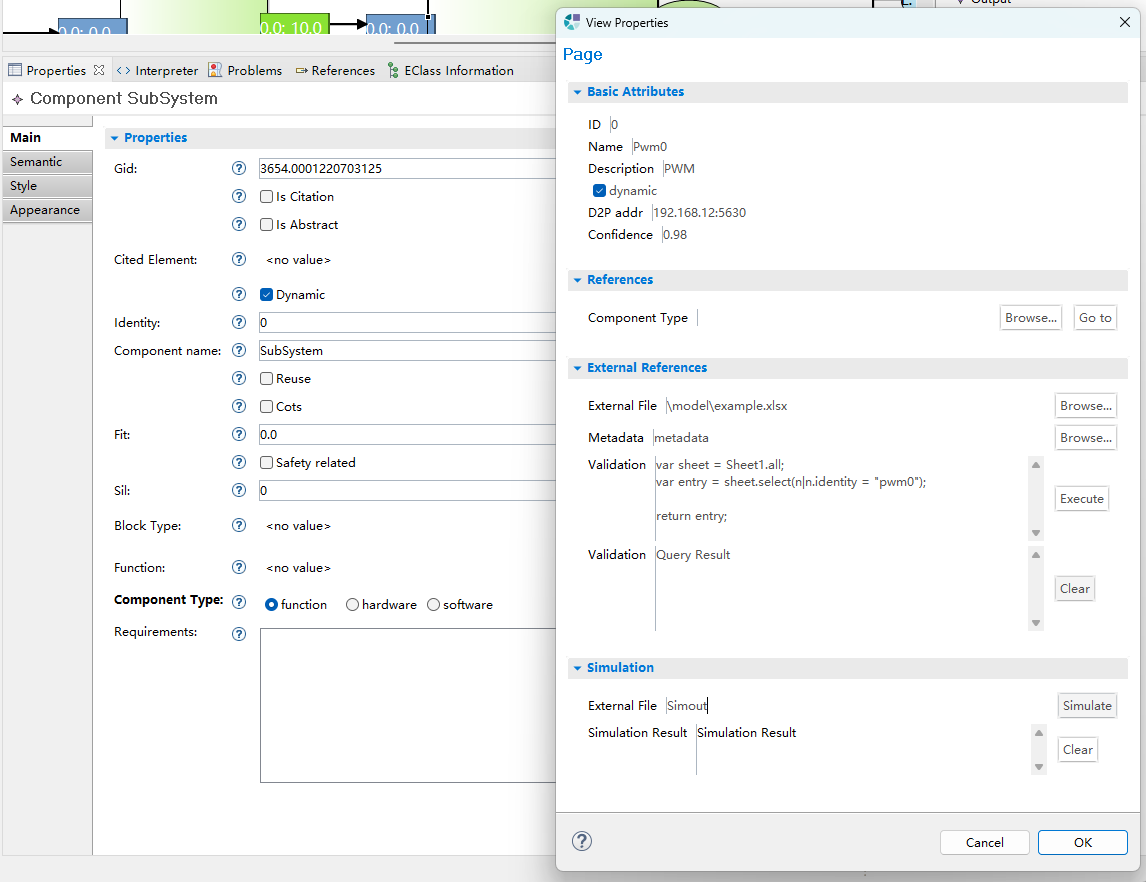}
	\caption{Trace to external heterogeneous models.} 
	\label{fig:property}
\end{figure*}

\begin{figure}[htb]
    \centering
    \includegraphics[width=\linewidth]{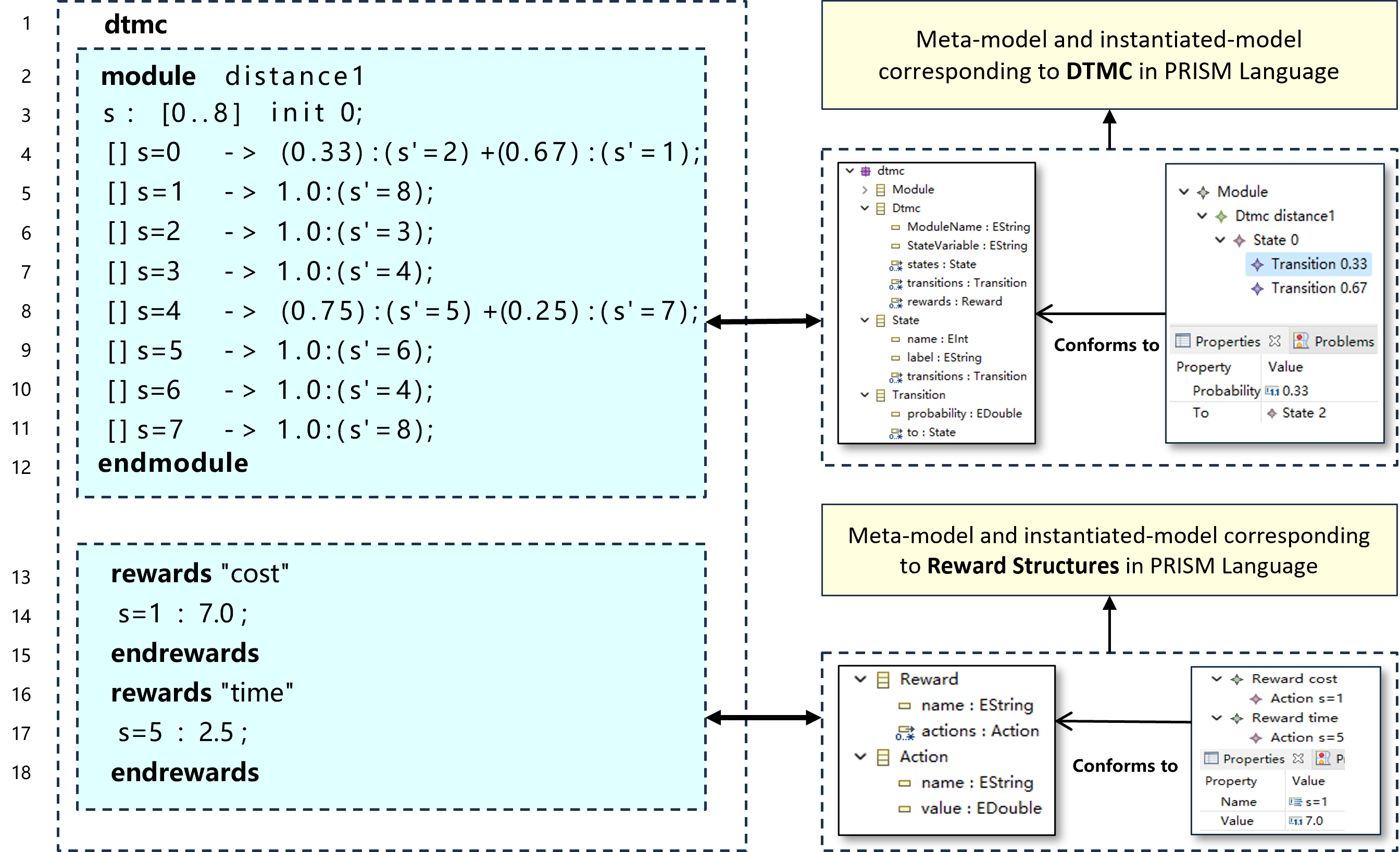}
    \caption{Discrete Time Markov Chain (DTMC) model.}
    \label{fig:dtmc}
\end{figure}

\begin{figure}[h!]
    \centering
    \includegraphics[width=1\linewidth]{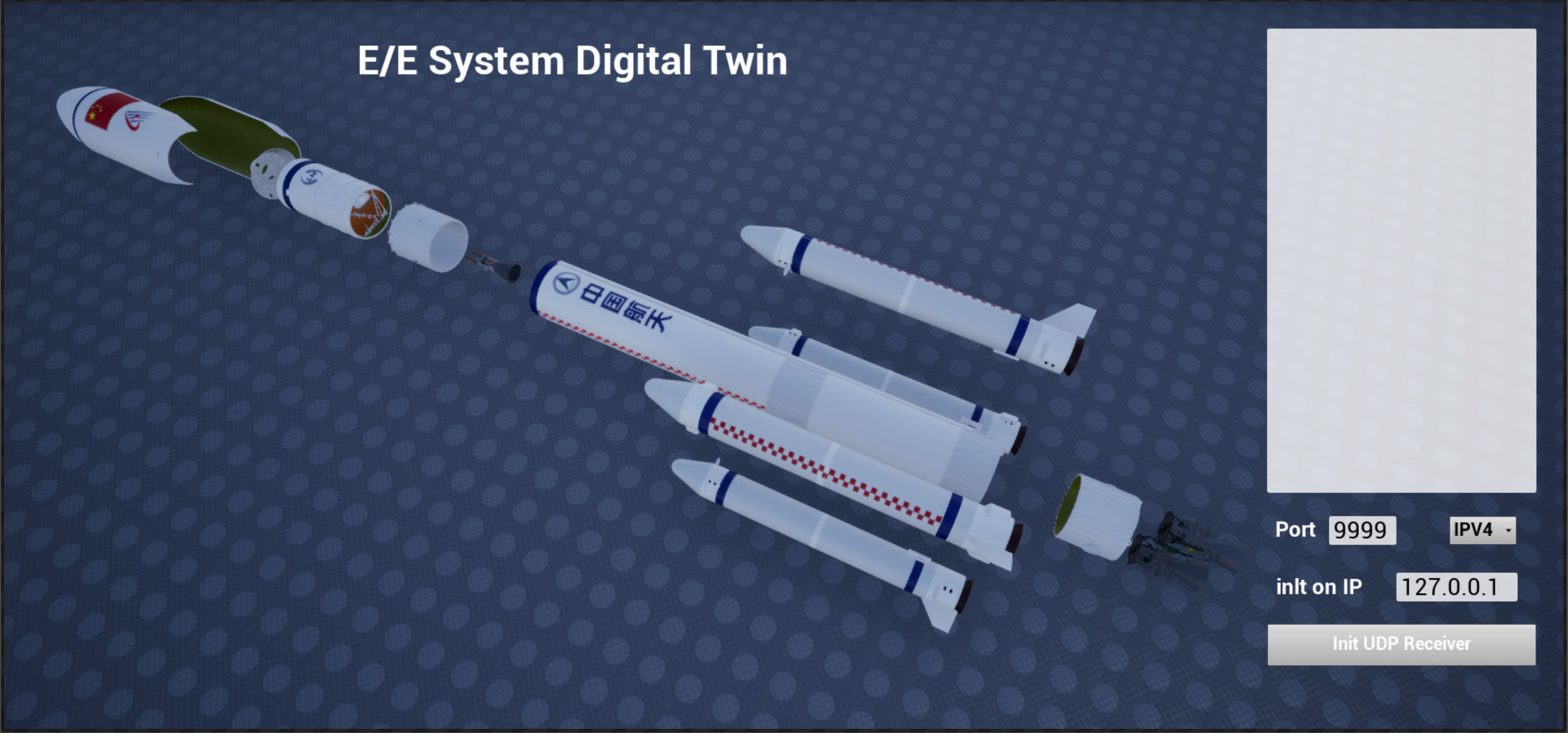}
	\caption{3D modelling of the space launch vehicle.} 
	\label{fig:3dmodel}
\end{figure}

\begin{figure}[h!]
    \centering
	\includegraphics[width=0.7\linewidth]{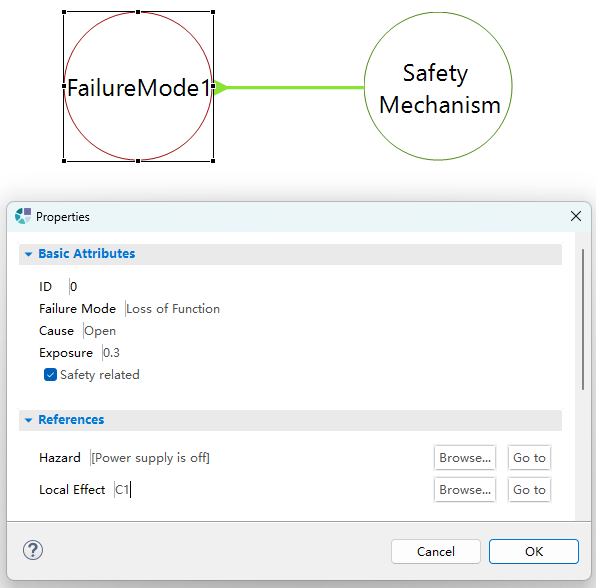}
	\caption{Failure mode modelling.} 
	\label{fig:failure}
\end{figure}

\subsection{DEVOTION Step 2: Enhancement of SDTM Models}
\subsubsection{Traceability links to heterogeneous models}
With DTME in place, together with the graphical modelling environment in place, in DEVOTION Step 2, we enrich the DTME model (i.e. the DM) by establishing traceability links from the model elements within the DTME model to heterogeneous models. 
For our case study, we want to establish traceability links to:
\begin{itemize}
    \item Requirement models, which are machine-checkable requirements for the reliability of the system.
    \item Simulink models, as shown in the upper part of Figure~\ref{fig:m2m}, which can be executed with dynamic configurations of the components.
    \item Reliability models, which contain the data regarding the reliability of components of the systems (e.g. their failure probability of failure and their failure modes), such information would be used by DTME to perform analysis on the system.
    \item Probabilistic models, in our use case, the Discrete Time Markov Chain (DTMC) model (as shown in Figure~\ref{fig:dtmc}), which will be updated with data from PT to perform analysis on the confidence of the components of the system.
    \item 3D geometric models (with their textures), for the E/E systems and the whole space launch vehicle, which can be used to assemble a virtualisation environment, as shown in Figure~\ref{fig:3dmodel}.
\end{itemize}

To establish the traceability links, we make use of the facilities provided in the \textit{Base} package of SDTM, namely the ``ExternalReference'' model element (in Figure~\ref{fig:base}). 
In the graphical modelling environment, we add the function to establish the traceability links, as shown in Figure~\ref{fig:property}. In the property view of each Component in SDTM, in the ``External References'' section, the users are able to specify the location of the model that they want to refer, its metadata, and also an executable script (written in EOL) which can be used to either obtain information from the model or validate the model. When the ``Execute'' button is pressed, the graphical editor will invoke Eclipse Epsilon and execute the script.

\subsubsection{Preliminary analysis on the enhanced DM}
With the enhanced DM that contains links to external models, we can specify the analytical algorithms to be performed on the DM for its reliability.
For illustration, we perform the \textit{Failure Mode and Effects Analysis} (FMEA).
FMEA is an analysis technique to analyse the effects of component failures to identify single point failures and calculate the metrics of system design to measure the integrity of systems. 
In order to perform FMEA, the data regarding component failure rates, component failure modes and the probability distribution of the failure modes is typically needed.

In our work, there are two approaches to perform FMEA.
In the first approach, we make use of the Simulink model and an external reliability model (in the form of an Excel spreadsheet), which contains: a) for each component type, their probability of failure; b) for each component type, their failure modes (e.g. for a Resistor, the failure modes can be \textit{open} and \textit{short}) and the distribution of the failure modes (e.g. if a Resistor fails, there is a 30\% chance it is caused by a \textit{open} and 70\% chance caused by a \textit{short}).
With the Simulink model and the reliability model in place, which are referenced from the SDTM model, we perform FMEA in an automated manner by calling the simulate function in Simulink, and ``inject'' failures by enumerating the failure modes in the reliability model, which is discussed in~\cite{wei2022designing}. 
In this way, we demonstrate that how simulation environment can be integrated with the DT platform and how simulations can help perform analysis on our DT model.

In the second approach, we make use of the facilities provided by SDTM and create the failure modes components, as shown in Figure~\ref{fig:failure}, we are able to create failure modes as well as safety mechanisms which are deployed to cover the failure mode, to significantly improve the reliability of the components.
With the second approach, we are able to record the reliability data of a component within SDTM, which is typically referred to as failure modelling.
With the failure logic embedded in SDTM, we can perform FMEA in an automated manner by using graph algorithms, which is discussed in~\cite{wei2023decisive}, and is not the focus of this work.

With FMEA, quantitative results can be obtained, a typical one is the Single Point Failure Metric (SPFM) specified in~\cite{iso26262}, which can be used to determine if the system is acceptably reliable. As previously mentioned, from the SDTM model, we also provide a traceability to requirement models. In our work, part of the requirement models are the non-functional requirements for the reliability of components, which are expressed as requiring the SPFM of components to a certain level. With the automated FMEA, we are able to check if the components satisfy their requirements, in an automated manner, which we will not go into details to aid readability.

\subsubsection{Version control and model indexing}
With the SDTM model containing references to external heterogeneous models discussed above, the next step is to put all of the models in a version controlled repository. For this purpose, we put all the models in a Git repository and monitor it with Eclipse Hawk. 
As previous discussed, Eclipse Hawk is a model indexing framework, which automatically indexes all of the contents in the Git repository in its graph database backend (currently we are using Neo4J as Hawk's backend). Since most of the models we are using are also EMF based, not much modification of Hawk is needed, except we need to develop Hawk's indexing algorithm so that it can understand the \textit{ExternalReference} model elements in the SDTM model and create edges within the backend (in Hawk's term, the \textit{derived properties}) so that corresponding information can be obtained through the edges when querying Hawk.

\subsubsection{Discussion}
Up to this point, we reached Maturity Level 2 of our DT by having a enhanced version of the DM. It is very crucial for the DM to integration information contained in different models in order to form an information graph so that analysis can be performed on the graph as whole. In addition, to promote scalability and enable version control, we used Eclipse Hawk for the indexing of the DM, as it would provide efficiency access to the information in the DM, in addition, it provides the functionality of querying the DM with timestamps, obtaining time-series data.

\subsection{DEVOTION Step 3: Enhancement DM with P2D and simulations}
\subsubsection{Generation of P2D through model transformations}
In this step, we focus on the generation of source code which enable the data flow from the PT to the DM.
For this purpose, we create a model-to-text transformation which create Java classes to receive message through the TCP protocol. 
This is typically done via declaring \textit{Component}s as \textit{dynamic} in the graphical editor of DTME.
In Figure~\ref{fig:property}, when a property dialog is opened for a \textit{Component}, the users may specify if the \textit{Component} is \textit{dynamic}.
After all \textit{Component}s are declared, if the user select \textit{Generate D2P} in the context menu of DTME, as shown in Figure~\ref{fig:dtme}, then a M2T is automatically executed, iterating all \textit{Component}s, collecting the ones that are declared as \textit{dynamic}, and generate a Java class, used to handle messages sent through TCP protocol, parse the message, and update the value of I/O nodes in the \textit{Component}s accordingly.
It is to be noted that the changes of values in the \textit{Component}s first take effect on the DTME model in the Git repository monitored by Hawk, Hawk then detects the change and index the change based on the timestamp of the change. The users of DTME may select to query Hawk using timestamps, and obtain a collection of values for \textit{Component}s as a time series function.

\subsubsection{Simulation based on current and historical data}
With the P2D facilities in place, in this step, we can perform simulations using the current and historical data. By doing this, we are able to identify the uncertainties that exist between simulation and the actual system, and may choose to compensate the uncertainties in the simulation to infer if there will be potential problems for the actual system.

As previously discussed, the DM contains traceability links to Simulink models, to invoke the simulation function, upon checking a \textit{Component}, as shown in Figure~\ref{fig:property}, the user may click on the ``Simulate'' button, DTME will trace to the Simulink model, locate relevant \textit{Component}, update the values of the blocks in the Simulink models, and invoke Simulink's simulate function. The output of simulation are typically a collection of time series values, which can be viewed within DTME.

\subsubsection{Discussion}
Up to this point, we reached Maturity Level 3 of our DT by transforming the DM into a DS by having generated P2D data flow, and by having capabilities to perform simulations of the system by integrating with simulation software. It is to be noted that within our work, we use Simulink as our simulation environment, we do not restrict the simulation environment, as long as they can be integrated into DTME.

\subsection{DEVOTION Step 4: Enhancement DS with D2P and visualisations}
\subsubsection{Generation of D2P through model transformations}
In this step, we focus on the generation of source code which enable the data flow from the DS to the PT.
For this purpose, we also leverage model-to-text transformations which create Java classes to send message through TCP. 
This is also done via declaring \textit{Component}s as \textit{dynamic} in the graphical editor of DTME.
In Figure~\ref{fig:property}, when a property dialog is opened for a \textit{Component}, the users may specify if the \textit{Component} is \textit{dynamic}.
In addition, for D2P, an IP address (with port number) is also needed as shown in Figure~\ref{fig:property}.
After all \textit{Component}s are declared, if the user select \textit{Generate D2P} in the context menu of DTME, as shown in Figure~\ref{fig:dtme}, then a M2T is automatically executed, iterating all \textit{Component}s, collecting the ones that are declared as \textit{dynamic}, and generate a Java class, used to send messages to the PT.
It is to be noted that currently we do not support the alteration of values of the E/E systems of the PT, when a value is changed in the DS, the D2P sends a message, which shuts down the PT.

\begin{figure}[h!]
    \centering
    \includegraphics[width=1\linewidth]{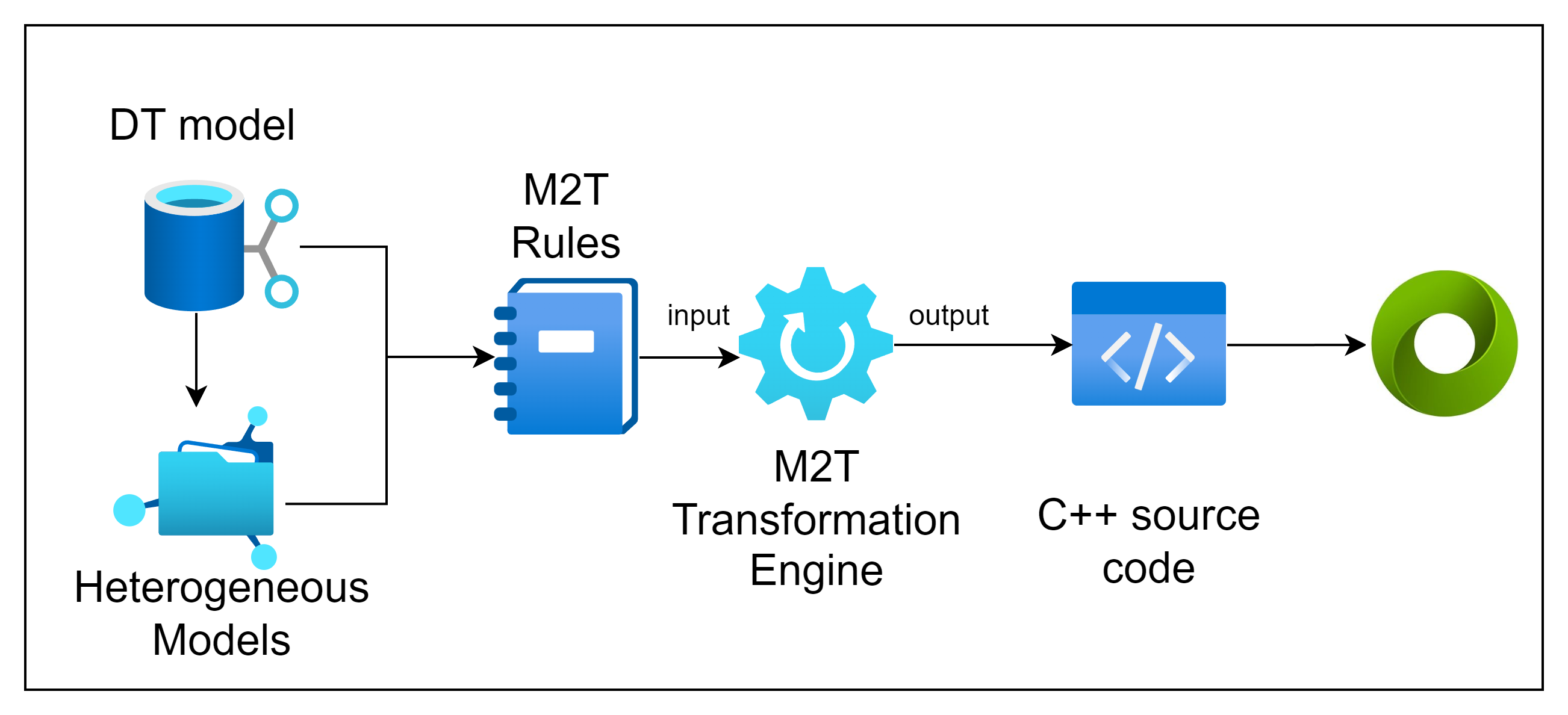}
	\caption{The generation of 3D virtual environments using MDE.} 
	\label{fig:virtual}
\end{figure}

\begin{figure}[h]
    \centering
    \includegraphics[width=1\linewidth]{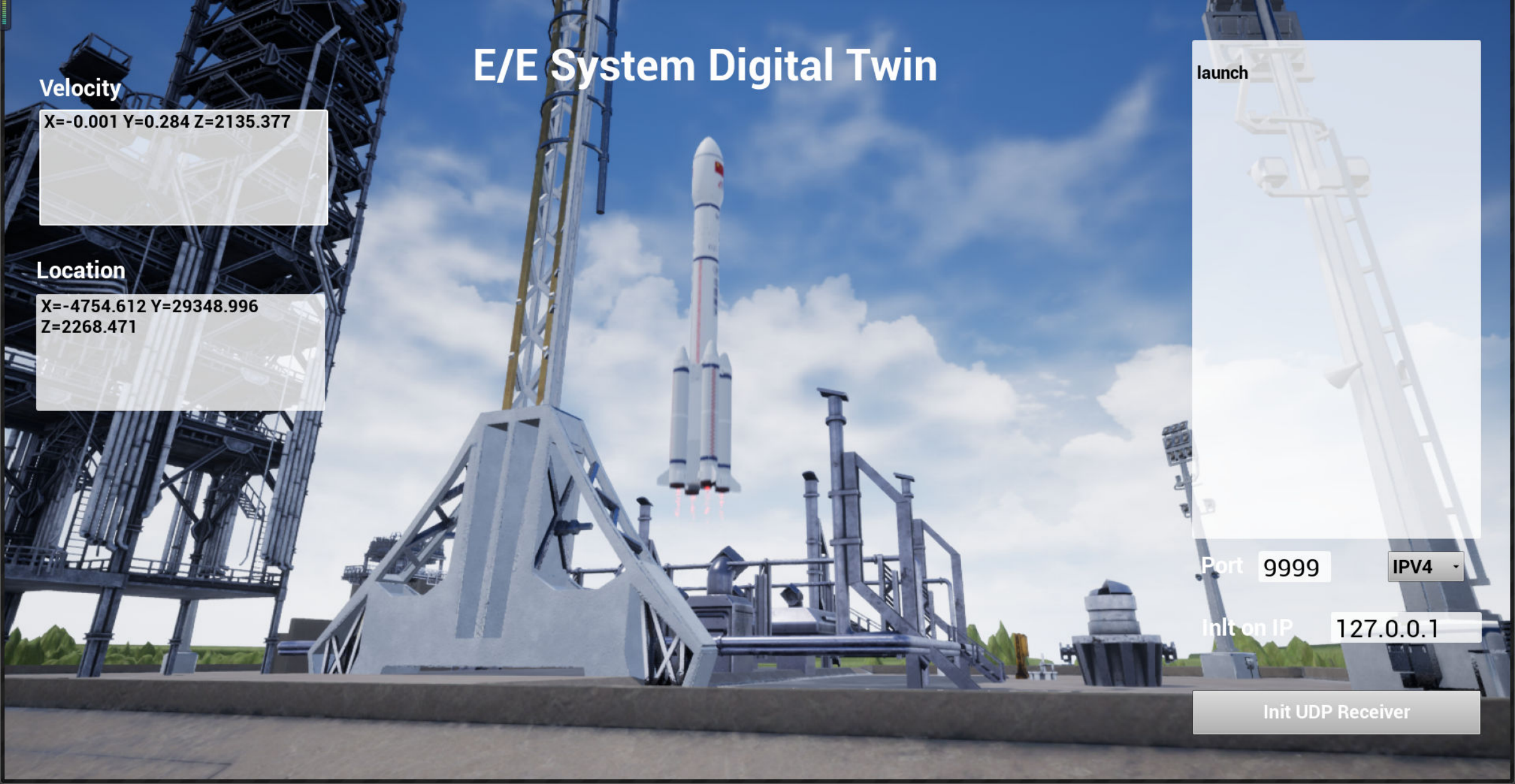}
	\caption{Rocket launch simulation, with DT and simulator synchronisation.} 
	\label{fig:launch}
\end{figure}
\subsubsection{3D based visualisation}
Up until now, the SDTM model evolves from a DS into a DT. 
However, it is not very feasible to test the DT as there are only a limited number of launches for space missions. 
Therefore, we rely on simulation in virtual environments created using Unreal Engine.
We touch upon briefly the misconception of DT -- that most perceive the DT as purely a simulation model created using 3D engines.
We would argue that 3D simulation is merely a means for visualisation, the real power of DT relies on the DT model and the operations applied on the DT models in order to gain insights and make decisions.

For the 3D virtual environment, we also adopt a model-based approach, as shown in Figure~\ref{fig:virtual}. 
We take the SDTM model (which contains traceability to external models, including 3D models) as input, and create a series of M2T rules, which are then executed by M2T transformation engine provided by Eclipse Epsilon, to generate C++ source code, which can be used to construct a 3D environment in the Unreal Engine (Figure~\ref{fig:3dmodel}).

To synchronise the DT with the Simulation, an independent data handling service is created, the D2P generation used in DTME communicates with this service and synchronise data with the 3D simulation in Unreal Engine. 
In the 3D simulation, we create a rocket launch simulation program, the program drives the 3D launch vehicle from ignition to the delivery of its payload. 
During the flight, the simulation program generates data for the E/E system DT and synchronise the data with the DT.
The DT then performs real-time reliability analysis as discussed previously, to provide feedback to the 3D simulation, in the sense that any reliability problems detected by the DT will reflect to the 3D objects that map the E/E System (we do not show the details of this due to confidentiality). 

\subsubsection{Discussion}
Up to this point, we reached Maturity Level 4 of our DT by transforming the DS into a DT, with D2P data flow mechanism automatically generated using model-to-text transformations. It is to be noted that within our work, we use Unreal Engine as our visualisation. In the future, we plan to use HTML web pages as visualisations, which can also be generated through model-to-text transformations.

\begin{figure}[h]
    \centering
    \includegraphics[width=1\linewidth]{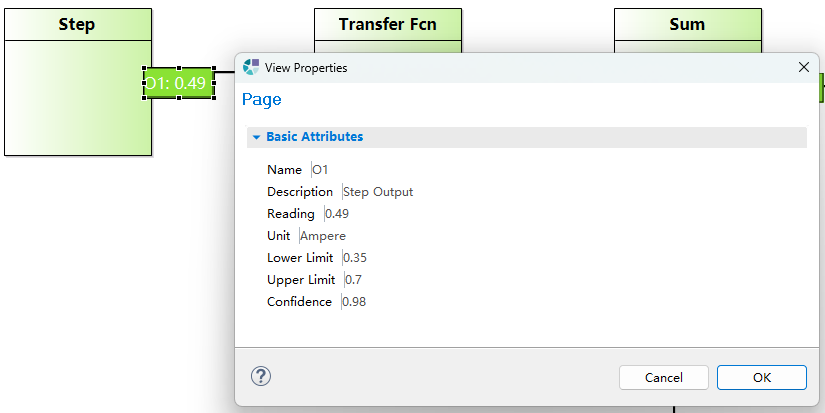}
	\caption{Component Output Modelling.} 
	\label{fig:ionodes}
\end{figure}

\subsection{DEVOTION Step 5: Enhancement DT with advanced algorithms}
\subsubsection{Probabilistic model checking to gain confidence of components}

With the a working DT in place, we briefly show how advanced algorithms can help with the reliability of the system using Model Checking (MC). 
MC is an effective means for developing systems and verifying the reliability of the system. 
Users can build a finite-state model and use MC to automatically verify whether the model satisfies requirements formally expressed in propositional temporal logic. 
MC can be utilised during the system design time to develop models that satisfy the system requirements and serve as a foundation for its implementation. Additionally, a variant of MC known as probabilistic model checking enables the verification of software systems against reliability, performance, and other quality-of-service (QoS) requirements. Probabilistic MC assesses models with state transitions annotated by probabilities or transition rates (e.g., discrete and continuous-time Markov chains), while operating with QoS requirements expressed in probabilistic versions of temporal logic.

For the purpose of gaining confidence for the system, we build a Discrete Time Markov Chain (DTMC) model, part of which is shown in Figure~\ref{fig:dtmc}. 
When the DT receives data through D2P data flow, the inputs and outputs of the components are updated. At the same time, the DTMC is updated checking if the values are within the legal ranges. Through multiple iterations, the DTMC would inform the confidence of each component, if the confidence level of a component falls below a threshold (currently set to 80\%), the DT would inform the PT to shutdown to prevent further failures. 

\subsection{Discussion}
Up to this point, we reached Maturity Level 5 of our DT by applying a probabilistic model checking algorithm to infer the health conditions of the system, and perform preventative shutdown if confidence decreases to a certain level. It is to be noted that we only report on our preliminary report, advanced algorithms such as data science and AI would be explored and applied in the future.

\section{Summary and Future Work}
\label{sec:conclusion}
In this paper, we performed a systematic review on the concepts of DT and the categorisations of DTs.
Based on the review, we proposed a DT maturity matrix, which systematically categorise different forms of models that describe physical entities and if they can be categorised as Digital Model (DM), Digital Shadow (DS) or Digital Twin (DT). 
From the DT maturity matrix, it can be seen that the development of DT is incremental. In order to develop a working DT, developers typically need to start with DM, gradually build facilities alongside the DM, evolve it into DS and then eventually into a DT.

Based on the above idea, we propose DEVOTION, a development methodology, using principles of Model Driven Engineering (MDE), to guide the development of DT through different stages of the DT maturity levels in the DT maturity matrix that we identified. 
We also provided a supporting tool, Digital Twin Management Environment (DTME) and a domain specific language, the Structured Digital Twin Metamodel (SDTM), which is highly modular, extensible, and can be used as the backbone to group all DT related processed data model to form a connected DT model. 
We discussed in detail the current packages in SDTM and we focused on how SDTM can be used to model the E/E System of space launch vehicles.

We then present a case study, in which we followed DEVOTION steps 1 -- 4. We discussed how MDE can facilitate a high degree of automation during the development DT and we showed how reliability analysis can be done using DT for both static and real-time data. We then touched briefly on the 3D simulation generation, and showed our preliminary results.

In the future, we will investigate on further development of DTME and achieve maturity level 5 for the space launch vehicle. We also aim to make the framework generic for any type of space launch vehicles. In addition, we are aiming to extend SDTM to capture other aspects of interest for space launch vehicles such as physics, mechanics, climate and improve the 3D simulation generation algorithms to achieve higher fidelity.

\balance

\bibliographystyle{abbrv}
\bibliography{references}

\end{document}